\documentclass[useAMS,usenatbib]{mn2e}

\usepackage{epsfig}

\newcommand{\vsini}{\mbox{$v_e\,\sin\,i$}}

\newcommand{\ha}{\hbox{$\hbox{H}\alpha$}}

\newcommand{\kmsec}{\,\mbox{$\mbox{km}\,\mbox{s}^{-1}$}}
\newcommand{\kms}{\,\mbox{$\mbox{km}\,\mbox{s}^{-1}$}}

\newcommand{\degs}{\mbox{$^\circ$}}

\newcommand{\speedy}{\hbox{Speedy Mic}}

\title[The X-ray corona of AB Dor]
      {The coronal structure of AB Dor determined from contemporaneous Doppler imaging and X-ray spectroscopy}
\author[G. A. J. Hussain, et al.]
{G. A. J. Hussain$^{1}$\thanks{E-mail: 
gajh@st-andrews.ac.uk; mmj@st-andrews.ac.uk; donati@ast.obs-mip.fr; bhouse@head.cfa.harvard.edu; 
 njd2@st-andrews.ac.uk; kw25@st-andrews.ac.uk; dupree@cfa.harvard.edu; acc4@st-andrews.ac.uk; Fabio.Favata@rssd.esa.int}, 
M. Jardine$^{1}$, J.-F. Donati$^{2}$, N. S. Brickhouse$^{3}$, N. J. Dunstone$^{1}$, 
\newauthor
K. Wood$^{1}$, A. K. Dupree$^{3}$, A. Collier Cameron$^{1}$, F. Favata$^{4}$ \\
$^{1}$ SUPA, School of Physics \& Astronomy, University of St Andrews, Fife KY16 9SS, Scotland. \\
$^{3}$ Harvard-Smithsonian Center for Astrophysics, 60 Garden Street, Cambridge, MA 02138.\\
$^{4}$ Astrophysics Division, Research and Science Support Department of ESA/ESTEC, Postbus 299, Noordwijk, Netherlands.
}
\begin{document}

\date{Accepted . Received ; in original form }

\pagerange{\pageref{firstpage}--\pageref{lastpage}} \pubyear{2006}

\maketitle

\label{firstpage}

\begin{abstract}
We obtain contemporaneous observations of the surface and corona of AB Dor, a young single cool star,  using ground-based circularly polarised spectra from the Anglo-Australian Telescope and X-ray lightcurves and spectra from the {\em Chandra} satellite. The ground-based data are used to construct surface magnetic field maps, which are extrapolated to produce detailed models of the quiescent corona. The X-ray data serve as a new test for the validity of these coronal models.

We find that AB Dor's X-ray corona must be  concentrated close to its surface, with a height, $H\sim0.3$--0.4R$_*$; this height is determined by the high coronal density and complex multi-polar magnetic field from the surface maps.  There is also significant correlation between the positions of surface and coronal active longitudes as determined from the surface spot and magnetic field maps and the X-ray lightcurve. At this epoch (2002 December)  AB Dor appears to possess one very large active longitude region, covering almost half the star; displaying enhanced activity in the form of large dark spots, strong magnetic fields and chromospheric emission. This is unusual as previous  surface maps of AB Dor typically display more active regions that span a wider range of longitudes.

Finally, the level of rotational modulation and shape of the X-ray lightcurve depend on the distribution of magnetic field in the obscured hemisphere (AB Dor is inclined by 60$^{\circ}$). The models that best reproduce the rotational modulation observed in the contemporaneous {\em Chandra} X-ray lightcurve and spectra require the magnetic field in the obscured hemisphere to be of the same polarity as that  in the observed hemisphere. The Sun shows different behaviour, with the leading polarity reversed in the opposite hemisphere. The X-ray observations provide a unique constraint on the magnetic structure in the obscured hemisphere. 
\end{abstract}


\section{Introduction}

Magnetic activity in rapidly rotating cool stars leaves distinctive signatures at all atmospheric levels: from stellar surfaces where strong magnetic fields cause the formation of large dark spots, to their coronae which typically have significantly larger temperatures, pressures and densities than those seen in the solar corona. 

Early X-ray missions, ROSAT and {\em Einstein}, established the global properties of X-ray emitting coronae in cool stars, revealing that all cool stars with outer convection zones possess hot MK coronae, with the rapid rotators displaying X-ray luminosities that are over 2 orders of magnitude larger than those observed on the Sun (Schmitt 1997). The latest X-ray and EUV missions such as EUVE, {\em Chandra} and XMM-{\em Newton} provide access to line diagnostics with high spectral resolution. This higher spectral resolution has been exploited to (a) probe the temperature distribution of hot, $>$ MK, plasma in active stellar coronae, revealing the distribution of coronal temperatures (e.g., Sanz Forcada, Brickhouse \& Dupree 2003; also see review by Dupree 2002), and to (b) measure coronal densities that are greater than $10^{10}$\,cm$^{-3}$ in the hot MK plasma (e.g.,Testa, Drake \& Peres 2004, Ness et al. 2002, G\"udel et al. 2001). These densities suggest typically small X-ray coronal structures (Dupree et al. 1993).

     While the physical properties of active stellar coronae are increasingly well determined, the detailed structure of stellar coronae is still a matter of debate; we have yet to ascertain how far the corona extends and how it is distributed around the star. This is important as the structure of the stellar coronal field affects how  zero-age main-sequence (ZAMS) stars spin down (Solanki et al. 1997), how binary systems interact, and even how planets form (Feigelson, Garmire \& Pravdo 2002). 
  Radio studies have found  that stable magnetic fields can exist out to several stellar radii around active RS CVn binary systems (e.g., Mutel et al. 1985). There is also evidence for very extended magnetic fields in the main sequence dM5.5e binary system, UV Cet A and B, 
from radio observations using the Very Large Baseline Array (VLBA). Images of  the strongest radio emitter appeared to have two stable radio-emitting  components caused by stable magnetic field structures at a distance of between 2.2 to 4\,R$_*$ (Benz, Conway \& G\"udel 1998). These sizes are consistent with the sizes of the stable prominences commonly detected as H$\alpha$ absorption transients on rapidly rotating cool stars (e.g., Collier Cameron \& Robinson 1989, Collier Cameron et al. 2002).

     While active stars are too distant to resolve directly, We can employ 
     indirect techniques to probe the structure of the hot, MK material in their coronae 
     using X-ray and EUV instrumentation. Recent studies of binary stars have enabled us to gain insights into where X-ray emitting regions are located: e.g., periodic variations in the EUV lightcurve of the contact binary 44i Boo strongly indicate that the  emission originates in a small, dense, high-latitude region on the primary star (Brickhouse \& Dupree 1998). Subsequent {\em Chandra} grating spectra confirmed this picture for 44i Boo using lightcurves and Doppler velocity measurements (Brickhouse, Dupree \& Young 2001). Hussain et al. (2005; hereafter, Paper I) have employed these techniques to the analysis of AB Dor. 
     
     X-ray flare analyses reveal the sizes of X-ray emitting coronal structures through eclipse  mapping techniques; e.g., a flare observed in the eclipsing binary Algol (B8\,V+K2\,IV) was completely eclipsed, thus strongly constraining its location to 0.5\,R$_*$ above the ``south'' pole of the K star (Schmitt \& Favata 1999; Favata \& Schmitt 1999). While flares on active G and K-type main sequence stars appear to originate within 1\,R$_*$ above the stellar surface (e.g., Mullan et al. 2006), a recent {\em Chandra} study has found that  young (1-10\,Myr) stellar objects in the Orion Nebula Cluster have flaring loops that can extend to much greater heights, up to 5--10\,R$_*$ (Favata et al. 2005).
     
     AB Doradus (AB Dor, HD 36705) is an active young (50\,Myr)  K0 dwarf that has recently arrived on the main sequence. It rotates very rapidly (Prot = 0.51479 day, v$_{\rm e} \sin i$ = 90\,km\,s$^{-1}$), so two rotation cycles can be covered in just over a day.  AB Dor is also bright at both optical and X-ray wavelengths (m$_V \sim $7,  $\log$\,L$_{\rm X}$/L$_{\rm bol}\sim30$ (Vilhu \& Linsky 1987). AB Dor displays signatures of magnetic activity at all atmospheric levels: Doppler maps have revealed the presence of large dark surface spots,  with a large spot covering its pole and smaller spots at lower latitudes, and strong kG field covering almost all observable latitudes (e.g., Donati et al. 1999, 2003); while long-term X-ray monitoring indicates that AB Dor flares almost once a day ({K\"urster} et al. 1997). AB Dor's emission measure distribution (EMD)  is typical for an active cool star, whether an RSCVn binary or a ZAMS star (Sanz-Forcada et al. 2002, 2003). Indeed, AB Dor displays characteristic magnetic activity signatures that appear typical of rapidly rotating cool stars in general.
     
 There is evidence for both compact and extended structures in the corona of AB Dor.
For example, small X-ray flaring loops are detected in flare decay analyses of X-ray lightcurves obtained using BEPPOSAX (Maggio et al. 2000).  They find that the flaring region is small ($H<0.3${\,\mbox{$\mbox{R}_*$}}) and that it
is uneclipsed over a full rotation cycle, indicating that it is located
near the pole of the star. A statistical analysis of ROSAT data spanning a
period of 5 years suggests a small level of X-ray rotational modulation at between 5-13\% ({K\"urster} et al. 1997). Brandt et al. (2001) find that between 60\% to 80\% of the UV
emission should originate near the stellar photosphere from HST/GHRS observations of AB Dor. 
 Radio emission from AB Dor shows rotational modulation; this modulation appears to be associated with magnetic structures lying over the most spotted regions, although the exact geometry  of the radio source depends strongly on the specific model used (Vilhu et al. 1993, Lim et al. 1994).

Conversely,  there is also  evidence for extended structures in AB Dor's corona, albeit from cooler material at transition region temperatures. For example, transition region lines observed in FUSE and HST datasets (C~{\sc iv}~1548\AA, Si~{\sc iv}~1393\AA\ and O~{\sc vi}~1032\AA) are observed to have broadened wings in emission extending to velocities of 270\,km/s. This broadening may be caused by optically thin plasma ($T \approx 10^5$~K) at heights out to the Keplerian co-rotation radius, 2.6{\,\mbox{$\mbox{R}_*$}}\ (Brandt et al. 2001, Ake et al. 2000).
The transition region line, O VI, shows modulation with rotational phase in both flux and profile.
Lower flux and narrower profiles correspond to the 
predicted locations of open magnetic field lines (Dupree et al. 2006). 
AB Dor also shows  evidence of extended structures, such as slingshot prominences
that extend out to over 2R$_*$; these are detected as fast-moving absorption transients in Balmer lines (e.g., H$\alpha$) (Collier Cameron \& Robinson 1989, Donati et al. 1999).
It is not clear whether or not they are in the open or closed corona although Jardine \& van Ballegooijen (2005) show how these structures can be trapped in long thin loops that are in the stellar wind, lying beyond the closed, X-ray emitting corona and extending out to 5\,R$_*$ in rapidly rotating stars like AB Dor.

We obtained contemporaneous observations of AB Dor at both optical and X-ray wavelengths in 2002 December in order to investigate the relationship between surface and coronal active regions. Surface spot and magnetic field maps have been obtained using the spectro-polarimetric optical data from the Anglo-Australian telescope (AAT) observations, and the stable, non-flaring X-ray emitting corona has been probed using the {\em Chandra} satellite. This is the first time that surface magnetic field and spot maps have been obtained contemporaneously with X-ray observations for an active cool star. We aim to extrapolate our surface magnetic field maps to construct detailed models of the X-ray emitting corona of AB Dor at this epoch. We will use these models to predict the amount of X-ray rotational modulation that we expect to see in this star assuming an isothermal corona and compare our predictions to the {\em Chandra} X-ray observations.

\section{Multi-wavelength campaign of AB Dor in 2002 December}

The active K0V star, AB Dor, was observed contemporaneously using the low energy transmission grating (LETG) and the high resolution camera spectroscopic array (HRC-S) on the {\em Chandra} X-ray satellite, and the SemelPol polarimeter on the UCL echelle spectrograph (UCLES) on the Anglo-Australian Telescope (AAT). The X-ray observations lasted almost two full rotation cycles (commencing at 21:25 UT on 2002 December 10 and  ending at 21:53 UT on 2002 December 11). No large flares were observed over this period; by covering almost two rotation cycles we could study X-ray rotational modulation caused by the quiescent non-flaring corona. 
The ground-based AAT observations spanned five consecutive nights (2002 December 11--December 15), overlapping with the X-ray observations for eight hours. Surface magnetic fields and spots on AB Dor remain stable  on timescales of over 1 week (e.g., Donati et al. 1997). Hence no significant changes will have occurred in the surface spots and magnetic field distributions over the course of the {\em Chandra} X-ray observations. 

The  high resolution spectro-polarimetry obtained from the SemelPol/UCLES/AAT configuration is 
inverted to reconstruct surface spot and magnetic field maps of AB Dor. Doppler imaging techniques were used to invert the AAT spectra and reconstruct detailed surface spot and  magnetic field maps of AB Dor (see Donati et al. 2003, 1999).   Full phase coverage was obtained over the observing run, resulting in surface maps with a latitude resolution of approximately 3$^{\circ}$. 

\subsection{{\em Chandra} X-ray observations: main findings}

The {\em Chandra} X-ray observations are described in detail in Paper 1.
Briefly, AB Dor was observed over  one continuous pointing spanning 88.1\,ksec 
(Observation ID 3762), which corresponds to 1.98\,$P_{\rm rot}$. 
Paper 1 reports  detailed evidence of rotational modulation observed in both the X-ray lightcurves and spectra; the main findings from Paper 1 are summarised  in Figure 1 and listed below: \\
\begin{itemize}
\item There is significant rotational modulation in the X-ray lightcurve, which has a flat level of emission superimposed with three peaks that cause 12\% 
rotational modulation.
\item The O VIII 18\AA\ line profile shows a rotationally modulated velocity pattern that repeats in two consecutive rotation cycles. The best-fit sine curve has a semi-amplitude of 30 $\pm$ 10 km s$^{-1}$.
\item The Chandra HETG Fe XVII 15\AA\ line profiles indicate that the X-ray emitting corona does not extend beyond 0.75\,R$_*$ above the surface, assuming that Doppler broadening is the cause of any excess broadening in the spectral line profile. We note also that  FUSE observations of the coronal forbidden Fe XVIII and Fe XIX lines at 974 and 1118\,\AA\ suggest an even smaller upper limit of approximately 0.5\,R$_*$.
\end{itemize}

\begin{figure*}
\epsfig{file=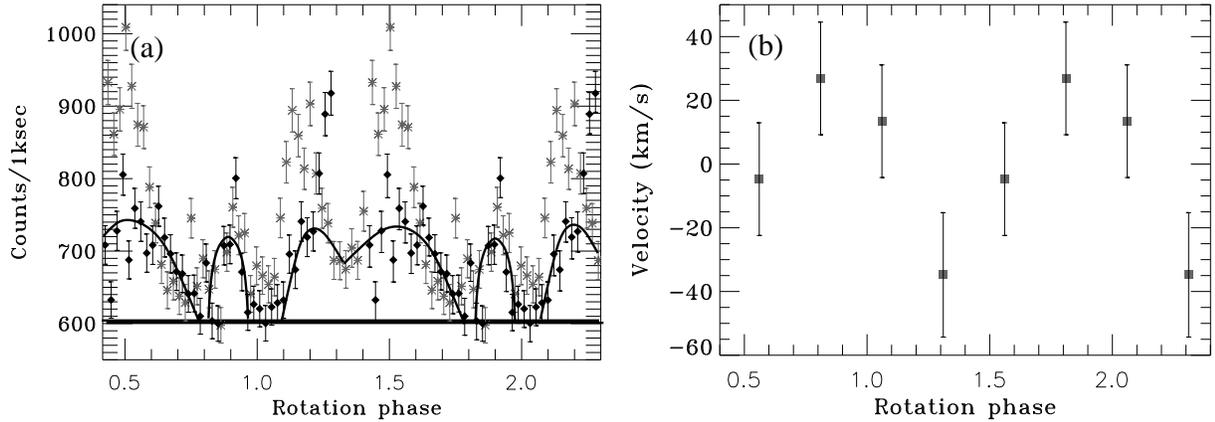, width=16cm}
\caption{(a) The X-ray lightcurve shows significant rotational modulation at the $\pm$12\% level. The lightcurve shows three components of emission: (i) a flat unmodulated level of emission (solid horizontal line), (ii) three peaks  (emphasised here by outlining the peaks with thick solid lines) that appear to repeat in the two rotation cycles and thus are rotationally modulated, and (iii) intrinsic flaring variability on short timescales. (b) The centroid of the OVIII 18\AA\ line profile traces ($\pm$30\,\kmsec) velocity shifts that appear to display rotational modulation over two consecutive rotation cycles. See Paper 1 for more details.}
\label{fig:recap}
\end{figure*}

\subsection{Ground-based AAT observations}

The SemelPol/UCLES/AAT configuration, used to obtain the ground-based spectro-polarimetric data,
has been described in detail in previous papers
(Donati et al. 2003, Donati et al. 1999, Donati \& Cameron 1999).
Briefly, SemelPol is a visitor instrument, mounted at the Cassegrain focus of the 3.9-m AAT and coupled with the UCLES spectrograph. 
The CCD chip records two beams of opposite polarisation spanning a wavelength range of 
over 2400\AA, from 4374.7\AA\ (order 129) to  6816.5\AA\ (order 84). 
The observations of AB Dor were obtained between 
2002 December 11--15 (see Table 1 for details). 

The cross-correlation technique of least squares deconvolution (LSD) is used to sum
up the signal from over 2700 photospheric lines, hence exploiting the full wavelength coverage
achieved with the AAT/SEMEL instrument configuration (Donati et al. 1999). 
The reduced spectra had a typical peak S:N level of approximately 146 to 200, 
the typical multiplex gain in 
S:N was a factor of 40--41, and the output LSD S:N  ranges between 5895 and 8512\AA\ (also see Table 1).

\subsection{Surface magnetic activity maps of AB Dor}

\begin{figure*}
\epsfig{file=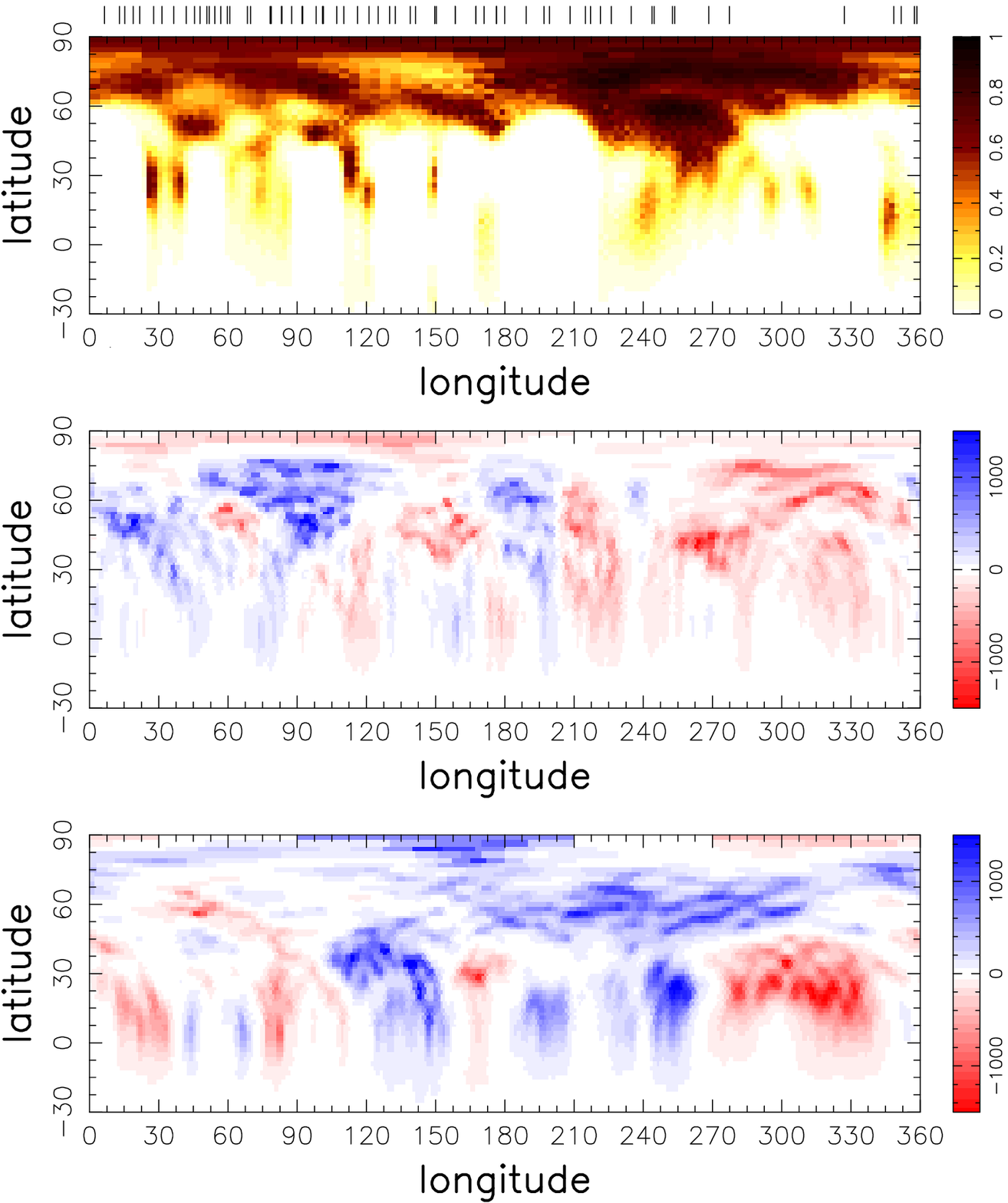, width=16cm}
\caption{Surface maps of AB Dor derived using SemelPol/UCLES/AAT spectra. These maps are rectangular projections of the stellar surface, latitudes below -30$^{\circ}$ are not observed as the star is inclined at 60$^{\circ}$. Observation phases are denoted as tick marks along the top plot. 
{\em Top:} Spot map of AB Dor from 2002 December; black denotes full spot coverage and white denotes unspotted photosphere. {\em Middle and Bottom:} Radial and azimuthal magnetic field maps of the stellar surface; blue and red denote +/- 1300\,G. 
These surface spot and magnetic field maps are characteristic of AB Dor; with strong field covering all observable latitudes and the presence of a large polar cap in the spot map. 
The fits to these data will be presented by Donati et al. (in prep.).}
\label{fig:maps}
\end{figure*}

Doppler imaging (DI) techniques enable us to produce  detailed surface spot and magnetic field maps of magnetic activity in rapidly rotating cool stars by inverting a time-series of high resolution spectra. The rotational broadening in the line profiles of active rapidly rotating cool stars separates the distortions caused by surface spots and magnetic fields in velocity-space
 (Vogt, Penrod \& Hatzes 1987,  Semel 1989, Collier Cameron \& Unruh 1994, also see review by Hussain 2004).
The DI code, DoTS, was used to reconstruct the spot map of AB Dor (Figure\,\ref{fig:maps}) by inverting a time-series of intensity line profiles (Collier Cameron 1997).
The technique of Zeeman Doppler imaging (ZDI) enables the reconstruction of surface magnetic field maps by inverting a time-series of  circularly polarised spectra from rapidly rotating stars. The code described by Donati \& Brown (1997) was used to produce the surface magnetic field maps shown in Figure\,\ref{fig:maps}. 

AB Dor's spot distribution has been mapped for over a decade. Its spot patterns are  typical of rapidly rotating active stars in general, with a large polar spot covering the polar region and extending to below 70$^{\circ}$ latitude. This polar spot appears to be 
almost unchanged for at least seven years. 
ZDI is a more recent technique, hence until recently AB Dor was the only active rapidly rotating star for which we had detailed magnetic field maps (latitude resolution $\sim 3^{\circ}$). The magnetic field maps in 2002 December are also very typical for AB Dor, with strong field covering almost all observable latitudes. Strong azimuthal field is found, with a strong unidirectional component at high latitudes; this is very similar to previous magnetic field maps of the star
(Donati \& Collier Cameron 1997; Donati et al. 1999; Donati et al. 2003).

\begin{table}
\hspace{-0.5cm}
 \centering
 \begin{minipage}{90mm}
  \caption{Observing log for 2002 December 11--15. The UT dates and times for each exposure are shown in the first three columns; $n_{\rm exp}$ is the number of intensity spectra/polarised spectra taken over the times listed; and t$_{\rm exp}$ is the the exposure time. The first row is the  {\em Chandra} observation, and the remaining rows are the  the AAT observations.}
\begin{tabular}{@{}lllll@{}}
\hline
Date	           &   MJD     & 	 UT (h:m:s)            & 	  n$_{\rm exp}$& 	t$_{\rm exp}$ \\
\hline
2002 Dec. 10--11 & 52618--9 & 21:25--21:53    &            & 88100 \\
\hline
2002 Dec. 11 &  52619.0  &  10:30:40/11:22:09 &    12/3 &   200  \\
&&	                                        11:50:53/12:41:24 &    12/3 &   200  \\
&&                                         13:13:19/14:22:16 &    16/4 &   200  \\
&&                                         14:49:58/15:59:01 &    16/4 &   200  \\
&&                                         16:28:46/17:37:41 &    16/4 &   200  \\
&&                                         18:07:54/18:27:52 &    4/1  &   200  \\
2002 Dec  12 & 52620.0  &   09:57:15/11:06:16 &    16/4 &   200  \\
&&                                         11:39:42/12:48:41 &    16/4 &   200  \\
&&                                         13:25:19/14:34:26 &    16/4 &   200  \\
&&                                         15:01:52/16:18:07 &    16/4 &   200  \\
&&                                         16:45:34/17:17:52 &    8/2  &   200  \\
&&                                         17:45:55/18:18:12 &    8/2  &   200  \\
2002 Dec 13  & 52621.0    & 09:45:13/10:35:51 &    12/3 &   200  \\
&&                                         11:02:54/11:53:34 &    12/3 &  200   \\
&&                                         12:22:15/13:12:52 &    12/3 &  200   \\
&&                                         13:39:23/14:30:02 &    12/3 &  200   \\
&&                                         14:57:26/15:48:06 &    12/3 &  200   \\
&&                                         16:22:31/17:13:05 &    12/3 &  200   \\
&&                                         17:40:35/18:12:44 &    8/2  &  200   \\
2002 Dec 14  & 52622.0    & 09:37:02/10:34:30 &    12/3 &  200   \\
&&                                         10:54:35/11:45:11 &    12/3 &  200   \\
&&                                         15:49:18/17:35:04 &    24/6 &  200   \\
&&                                         18:03:31/18:17:20 &    4/1  &  200   \\
2002 Dec 15  &  52623.0  &  11:08:20/11:40:36 &    12/3 &  200   \\
\hline
\end{tabular}
\end{minipage}
\end{table}

\subsection{Comparison of active longitudes at the surface and corona}

\begin{figure}
\hspace{-0.3cm}
\epsfig{file=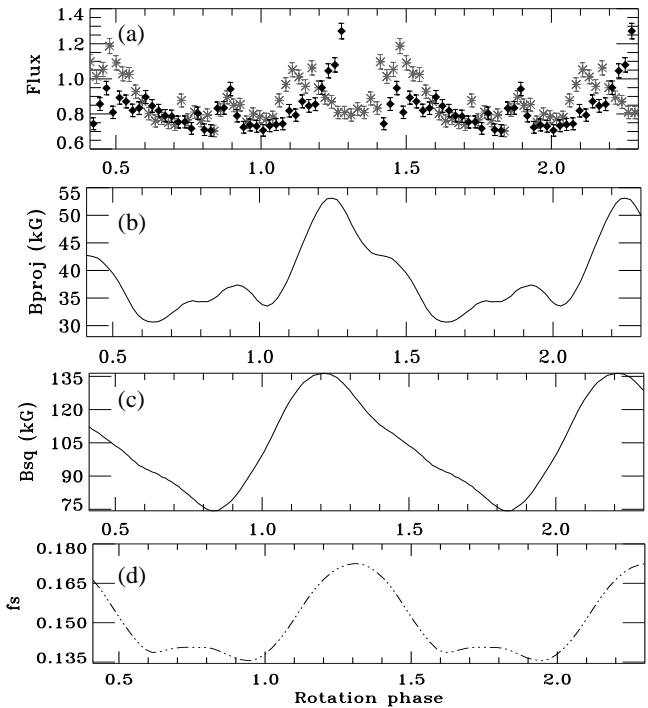, width=8.5cm}
 \caption{Comparison of active longitudes in coronal (X-ray) and surface (magnetic field and spots) active regions. (a) {\em Chandra} X-ray lightcurve -- asterisks and filled symbols represent the first and second rotation cycle respectively. The  lightcurve has been normalised so that the lower envelope of the X-ray emission, representing the component that is rotationally modulated lies below 1.0. X-ray emission above 1.0 is caused by short-term, most likely flaring, variability and is excluded from our calculations for the Kendall $\tau$-test. Figures (b)--(d) have been computed using the surface magnetic field and spot maps shown in Figure 2; (b) $B_{\rm proj}$ is the total magnetic field projected along the line-of-sight summed over each phase; (c) the total magnetic field squared, $B^2$ from the Zeeman Doppler maps shown in Figure 2; and (d) the spot coverage ($f_s$) computed from the spot map in Figure 2 as a function of rotation phase.
}
\label{fig:activelon}
\end{figure}

The data acquired here enable us to compare the longitude positions  of active regions at all atmospheric levels: from the surface (as probed by the surface spot and magnetic field maps) to the corona (as probed by the X-ray lightcurve). As discussed in Paper 1, the 
X-ray lightcurve can be described as a flat unmodulated emission level superimposed with the enhanced X-ray emission from two or three active regions; these active regions cause peaks in the lightcurve near phases 0.2, 0.5, and 0.9. The peaks also seem to be associated with short-lived energetic flaring regions. The lower envelope of emission repeats fairly well  in both cycles (particularly the peaks near phases 0.9 and 0.5 ), suggesting the presence of coronal X-ray emitting structures that remain stable for the period of observation. 
Figure\,\ref{fig:activelon}a shows the X-ray lightcurve: this has been normalised using the peak counts in the stable, rotationally modulating part of the X-ray lightcurve. We attribute the emission above 1.0 in this plot to emission from short-lived flaring regions. 

Figures\,\ref{fig:activelon}(b--d) have been computed using the surface spot and magnetic field maps in Figure\,\ref{fig:maps}. Spot coverage ($f_s$) ranges from 1.0 (representing spots covering the entire projected stellar disk) to 0.0 (representing unspotted photosphere). 
Figure\,\ref{fig:activelon}(d) was computed by projecting the surface spot map onto the inclined stellar disk at each rotation phase covering a full rotation cycle. 
The spot coverage at each rotation phase is then computed by integrating the spot coverage over the entire inclined stellar disk. A similar process was carried out with the surface magnetic field maps to compute $B^2$ and $B_{\rm proj}$ as a function of rotation phase. 
Figure\,\ref{fig:activelon}b shows the square of the line-of-sight component of the magnetic field ($B_{\rm proj}=|B\sin{\theta}|^2$) as a function of rotation phase.
Figures\,\ref{fig:activelon}b--d show the presence of a large active region at the stellar surface that peaks near rotation phase 1.3, very near the centre of the two largest peaks in the X-ray lightcurve (Fig\,\ref{fig:activelon}a). 

The significance  of this correspondence can be quantified using the non-parametric statistical Kendall $\tau$ test. 
This test measures the strength of dependence between two variables, quantifying the degree of correlation or lack of correlation. We investigate whether or not there is any correlation between the positions of surface spots and X-ray emission in  AB Dor.
As required by the Kendall test, both datasets need to have the same binning. 
We adopt the same time-bins used in the X-ray lightcurve when computing the $f_s$ and 
$|B_{\theta}^2|$ vs. phase plots;  the observation phases between 0.10 to 0.23 are omitted in the subsequent calculation as there are phase gaps in this region in the surface maps and therefore spot and magnetic field detections in these areas are less reliable. 
Similarly the flaring regions (above 1.0 in the top plot of Figure\,\ref{fig:activelon}) are omitted in these calculations as we are only interested in the quiescent coronal emission. 
The $\tau$ test is used to rank-order the $f_s$ and X-ray emission, and the $|B_{\theta}^2|$ and X-ray emission; it enables us to measure the degree of correlation, the statistical probability of this correlation being due to chance (Press et al. 1988).  We find highly significant correlation between all three quantities investigated: i.e. X-ray flux, spot coverage and $|B_{\theta}^2|$.

 \begin{table}
  \centering
 \begin{minipage}{140mm}
  \caption{$\tau$ test of magnetically active regions}
  \begin{tabular}{@{}llll@{}}
  \hline
                         & $\tau$ & Probability \\               
\hline
X-ray vs. $f_s$            & 0.276 &  5.10 10$^{-4}$\\
X-ray vs. $|B_{\theta}^2|$ & 0.308 &  1.03 10$^{-4}$\\
$|B_{\theta}^2|$ vs. $f_s$ & 0.500 &  2.80 10$^{-10}$\\
\hline
\end{tabular}
\end{minipage}
 \end{table}

\section{Building a coronal model of AB Dor}

Jardine et al. (2002) extrapolate surface magnetic field maps of AB Dor and use the resulting three-dimensional magnetic field models to predict coronal properties of rapidly rotating active stars (e.g., X-ray emission levels and rotational modulation,  locations of open field; Jardine et al, 2006). The contemporaneous AAT and {\em Chandra} datasets enable us to test our X-ray modelling capabilities in detail for the first time. 
The X-ray corona is modelled using the following inputs: (a) the three-dimensional magnetic field model extrapolated from the surface magnetic field map (Figure 3), (b) an isothermal corona that has a temperature of approximately 10\,MK (consistent with emission measure distributions of AB Dor), and (c) assuming hydrostatic equilibrium. The extent of the closed corona depends on the strength and complexity of the surface magnetic field maps, the emission measure level and the electron density of hte X-ray corona of the star (as determined from X-ray observations of AB Dor).  This dataset represents a unique opportunity to use measurements from X-ray data obtained contemporaneously with magnetic field maps to test our models and help in constructing more realistic models.

\subsection{Coronal fluxes and densities in 2002 December} 

The  {\em Chandra} data were re-reduced using the latest dedicated CIAO pipeline processing routines (CIAO v.3.3.0.1). The most prominent He-like triplet, the O\,VII triplet, is used to to measure AB Dor's coronal density in 2002 December. The resonance, intercombination and forbidden lines are all detected clearly in these X-ray spectra. Measurements based on previous observations (Testa, Drake \& Peres 2004, Sanz-Forcada, Maggio \& Micela 2003) can be compared with our measurements to estimate the stability of the densities in the X-ray emitting regions of the stellar corona.

\begin{figure}
\epsfig{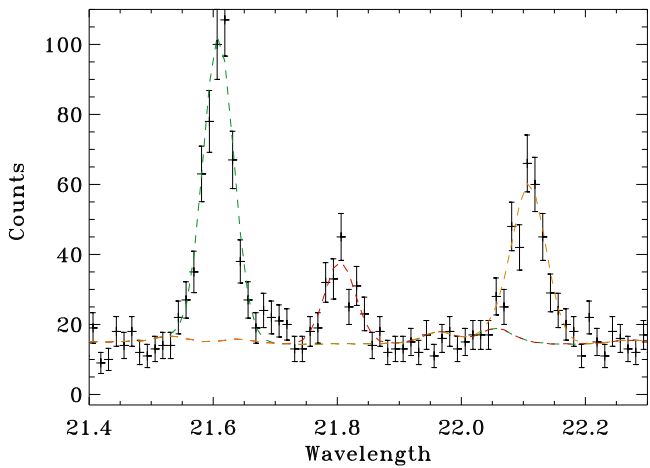}
 \caption{The He-like density-sensitive  OVII triplet at 21\AA.
}
\label{fig:density}
\end{figure}

In order to measure line fluxes for this triplet we first had to evaluate the continuum level accurately. We estimate the continuum level by generating fake  or model continuum and spectra using the EMD derived by Sanz-Forcada, Maggio \& Micela (2003) from {\em Chandra}/HETG/ACIS-S 
spectra of AB Dor obtained in 1999 October (Observation ID 16). 
The spectra are folded through the LETG instrument response function, modelling for combined orders up to order 5. We find that the fluxes from these model spectra are a good fit to the observed spectra with very little scaling required suggesting that the emission measure distributions are  very similar at both epochs (1999 October and 2002 December; Sanz-Forcada, Maggio \& Micela 2003). 
These model spectra are used to determine the best line-free regions, and the model continuum is scaled to fit these areas. As these regions are relatively small, there is still considerable uncertainty in the placing of the continuum level and this contributes the largest uncertainty to the subsequent flux measurements. 
Once the continuum level is determined,  each line profile is fitted using a least-squares Gaussian fitting procedure; the exact shape of the profile fit does not make much difference to the final measurements. 
We also fit the Moffat function, $I=(1+(\lambda/\lambda_c)^2)^{-\beta}$,  with $\beta=2.5$, as this shape is a closer representation of the instrumental line profile (Chandra Proposers' Observatory Guide Section 9.3.3 v.8),  however the resulting fluxes are consistent with the Gaussian fitting method. The measured fluxes are reported in Table\,3\, 
and the fits are shown in Figure\,\ref{fig:density}. 

The ratios: $f/i= 2.1\pm 0.2$ and $(f+i)/r=0.86 \pm 0.1$ correspond to an electron density, 
$\log$\,n$_e {\rm [cm^{-3}]} \sim\,10.3^{+0.4}_{-0.3}$. 
The $(f+i)/r$ ratio gives the electron temperature, and is found to be  2.2\,MK (using the Astrophysical Plasma Emission Database, APED; Smith et al. 2001). 
The errors quoted for the density take the  largest source of  uncertainty into account; i.e. the placement of the continuum level (Brickhouse 2002).
 This uncertainty dominates the measurement errors associated with line profile fitting. For instance, if the continuum is actually lower than the value we adopt (e.g., scaled by 80\% of the current value), this would increase the  measured $n_e$ by a factor of 2,  while still fitting the continuum level in the line-free regions of the X-ray dataset.

Measurements of densities at higher coronal temperatures are harder to determine with our 
LETG dataset as either the diagnostics lines are at short wavelengths where the spectral resolution is poor or else they are at very high wavelengths (91--128\AA) where the background level is high and the lines are relatively weak.

AB Dor's coronal $n_e$ has previously been estimated using the OVII He-like density sensitive ratio from the data obtained using MEG/ACIS-S and XMM-{\em Newton}/RGS observations of the star. 
 A full table of previous measurements of $n_e$ are shown in 
Table\,\ref{table:ne}.
These measurements show that our measured value of $n_e$ 
is consistent with previously measured values, especially given the error introduced by the difficulties  of measuring the continuum level accurately. 
The Ne\,IX and Mg\,XI diagnostics, formed at higher temperatures, show higher densities.
Using {\em Chandra}/HETG spectra, 
Sanz-Forcada et al. (2003) report  $\log n_e{\rm [cm^{-3}]}  \sim 11$
using Ne\,IX 
($T=3.98$\,MK) and  $\log n_e{\rm [cm^{-3}]} \geq 12$; Ness et al. (2004) measure
 $\log n_e[cm^{-3}]=11.4\pm 0.16$ using Ne\,IX ($T=4$\,MK); 
and Testa et al. (2004) measure $\log n_e{\rm [cm^{-3}]}  <12.25$ using Mg\,XI 
($T=3.98$\,MK).

 \begin{table*}
  \centering
 \begin{minipage}{140mm}
  \caption{Density measurements for the He-like OVII triplet.}
  \begin{tabular}{@{}lllll@{}}
  \hline
Density & Temperature & Source & Date & Reference \\
$\log n_e [cm^{-3}]$ & [MK] & & & \\
\hline
$10.78\pm 0.14$            & 1.6$^{0.9}_{-0.5}$   & MEG &1999 Oct  & Sanz-Forcada et al. 2003\\
$10.2 \pm 0.74$             & 2                          & MEG &1999 Oct  & Ness et al. 2004 \\
$10.75^{0.35}_{-0.12}$  & 2                          & MEG & 1999 Oct  & Testa et al. 2004 \\
$10.4^{0.32}_{-0.36}$	& 2   	    		         & RGS & 2000 May  & G\"udel et al. 2001\\
$10.5 \pm 0.10$             & 2                          & RGS & 2000 May & Ness et al. 2004\\
$10.77\pm 0.08$            & 1.7 $\pm$ 0.3	         & RGS & 2000 Jun  & Sanz-Forcada et al. 2003\\
$10.82\pm 0.06$            & 2.1$\pm$ 0.3   	   & RGS & 2001 Jan  & Sanz-Forcada et al. 2003\\
$10.3^{0.4}_{-0.3}$		& 2.2 		              & LETG & 2002 Dec & {\em this paper} \\
\hline
\end{tabular}
\end{minipage}
\label{table:ne}
 \end{table*}

 \begin{table}
  \centering
 \begin{minipage}{140mm}
  \caption{Flux measurements for the He-like OVII triplet.}
  \begin{tabular}{@{}lllll@{}}
  \hline
Wavelength (\AA)							& 21.6 (r) 	& 21.8 (i)	& 22.1 (f) 	& \\
Fluxes	 ($10^{-5}$)erg/cm$^2$/s	& 26.8    	& 7.4    	& 15.6	& \\
Errors   ($10^{-5}$)erg/cm$^2$/s	&    1.9 	& 1.3		&   1.6\\
\hline
\end{tabular}
\end{minipage}
\label{tab:fluxdensity}
 \end{table}
 
\subsection{Constructing models: reconciling field, EMD and $n_e$ measurements}

\begin{figure*}
\epsfig{file=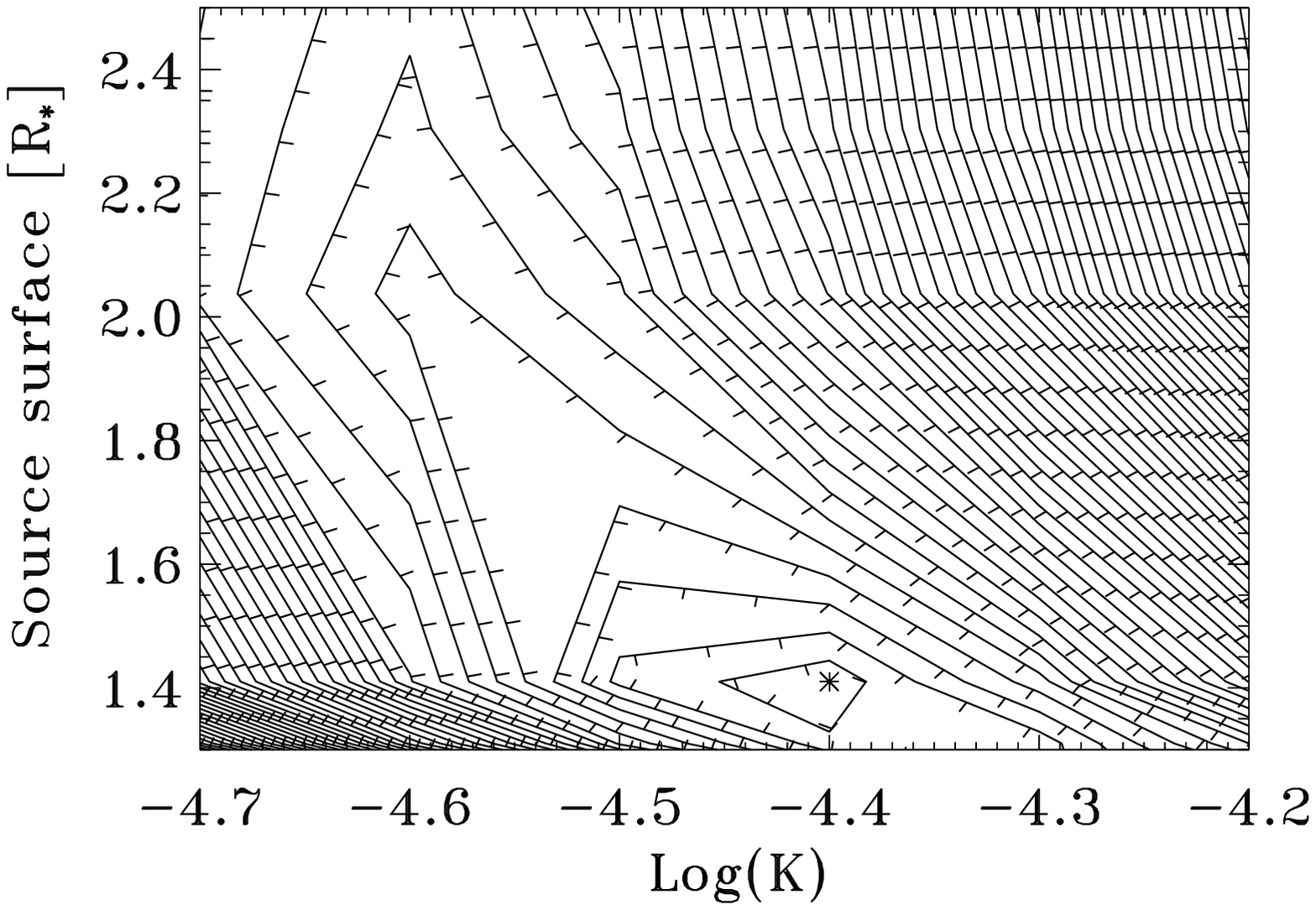, width=8.5cm}
\epsfig{file=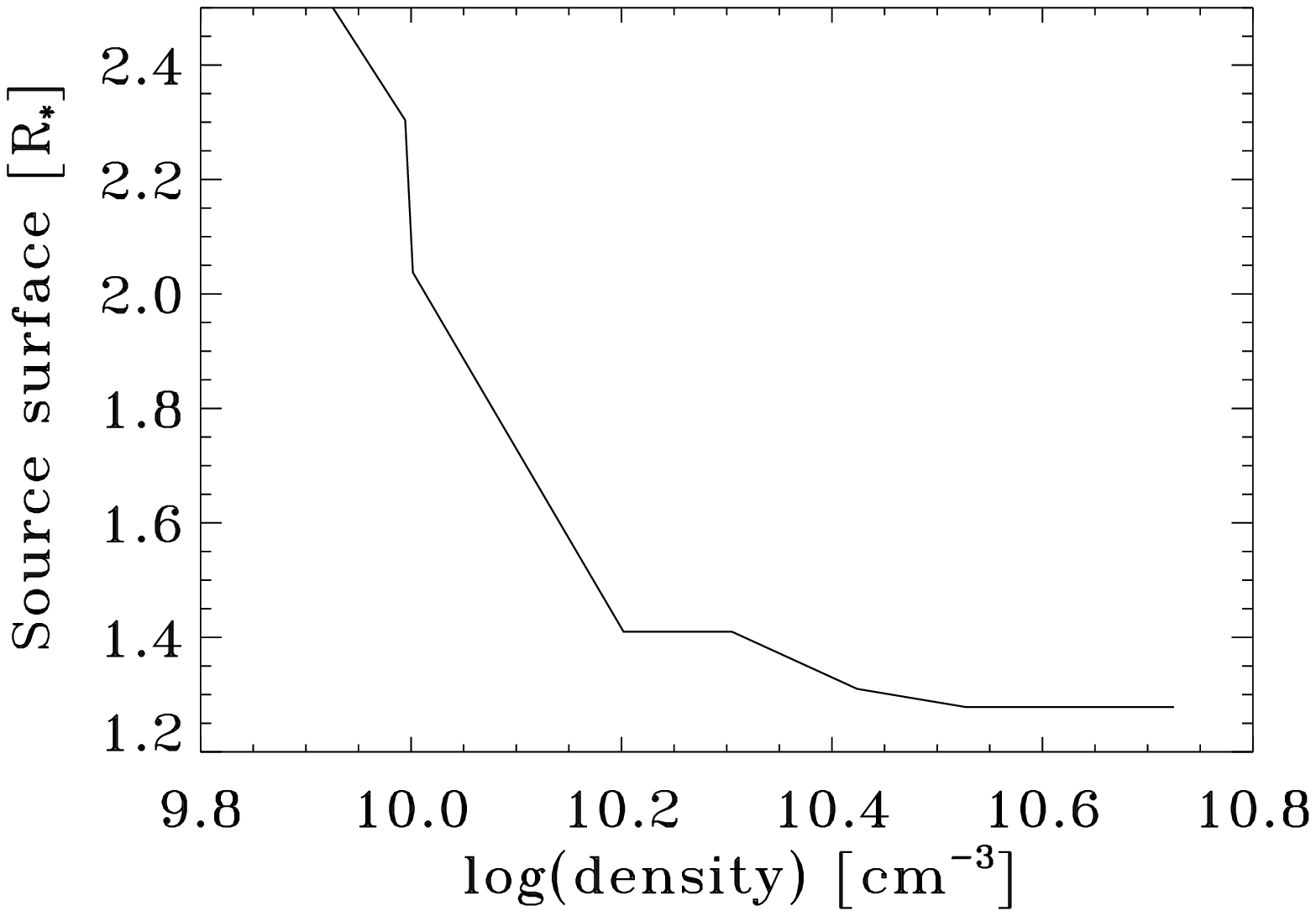,width=8.5cm} 
\caption{{\em Left: } Contours of goodness-of-fit that are used to select the best values of  the radius of the source surface, $R_{\rm ss}$ and, $K$, the scaling factor determining the pressure at the base of X-ray emitting loops. {\em Right: }
Higher densities necessitate a more compact corona. This plot shows the relationship between the  extent of the X-ray emitting corona and the coronal density, assuming the magnetic field map shown in Figure\,1 and an isothermal corona with T=10\,MK and a $\log{\rm (EM) [cm^{-3}]}=52$. Our measured  density, $\log n_e{\rm [cm^{-3}]}=10.3$, suggests a very compact corona with a source surface at a radius, $R_{\rm ss}\sim1.4$R$_*$}.
\label{fig:model_rss}
\end{figure*}

We build a simple isothermal coronal X-ray model by extrapolating the surface magnetic field maps in Figure\,\ref{fig:maps}. All of the closed field lines are assumed to be filled with isothermal X-ray emitting gas in hydrostatic equilibrium. 
We take the temperature of the corona to be $T\sim 10$\,MK in accordance with the stable peak in AB Dor's EMD (e.g., Sanz-Forcada et al. 2003). 
Before we compute the X-ray properties of the corona at this epoch 
we have to determine two parameters:
the radius of the source surface -- this determines the extent of the closed, X-ray emitting corona and is the point beyond which all the field becomes radial, and the pressure at the base of the X-ray emitting loops. The pressure at the footpoint of 
closed loops,  $p_0$, scales with the magnetic field strength, {\bf B}, by a factor K:
$p_0=KB^2_0$. 
  This scaling factor, $K$, and the extent of the closed, emitting corona depend on the following parameters: magnetic field strengths from the surface field maps,  coronal density, temperature and emission measure, EM. The X-ray values we are trying to fit here are:
coronal temperature ($T_c$=10\,MK), density ($\log n_e[cm^{-3}]=10.3$) and emission measure
${\log \rm EM[cm^{-3}]}=52$.

The coronal X-ray emissivity is assumed 
to be proportional to the density squared and we use a three-dimensional Monte Carlo  radiation transfer code to produce optically thin X-ray images (assuming no scattering or absorption of the coronal X-rays). 
 Figure\,\ref{fig:model_rss} shows contours of goodness-of-fit to the observed X-ray parameters (i.e. $T_c$, $n_e$ and EM) of our extrapolated magnetic field model as a function of the radius of the source surface, R$_{\rm SS}$, and pressure scaling factor, $K$.
As the coronal models are in hydrostatic equilibrium, the model densities are 
actually an emission-measure weighted average over the volume of the corona.
Figure\,\ref{fig:model_rss} shows the dependence of the extent of the X-ray emitting corona on the coronal density; unsurprisingly low densities are consistent with  more extended coronae. The extent of the corona can also be increased by increasing magnetic field strengths and/or increasing the EM value or temperature. However, given AB Dor's surface magnetic field map obtained using ZDI, 
and its coronal  EM, $n_e$ and temperature, our results suggest an X-ray emitting corona that is very compact with a source surface at radius, 
$R_{\rm ss}=1.4$\,R$_*$.

\section{Testing the coronal models}
\begin{figure*}
\epsfig{file=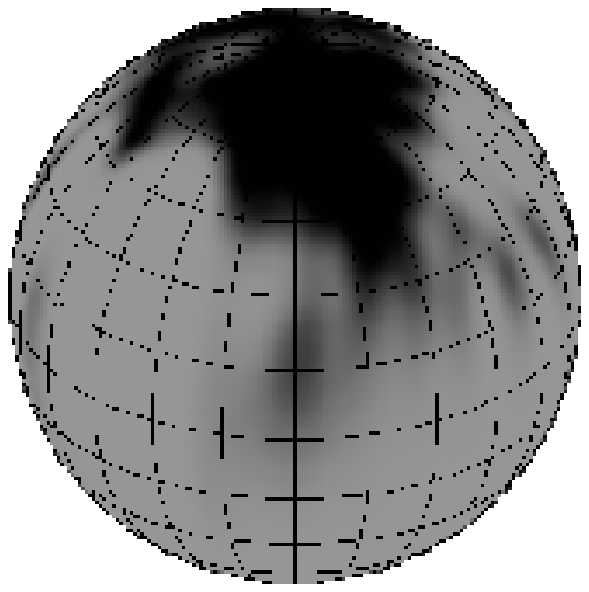, width=7cm}
\epsfig{file=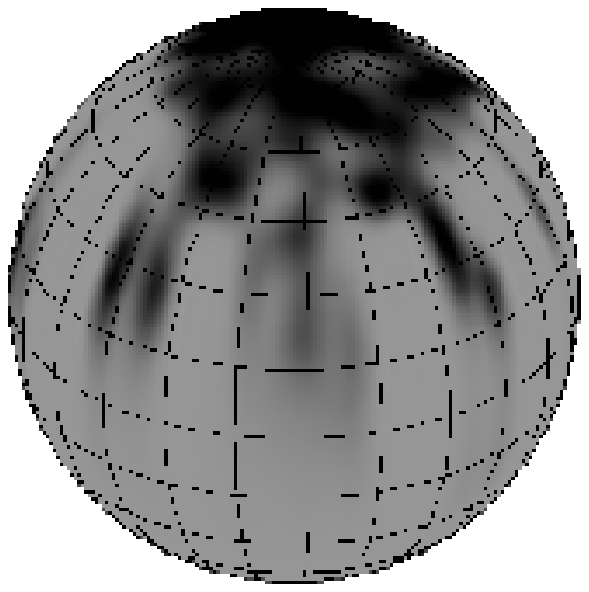, width=7cm}
\epsfig{file=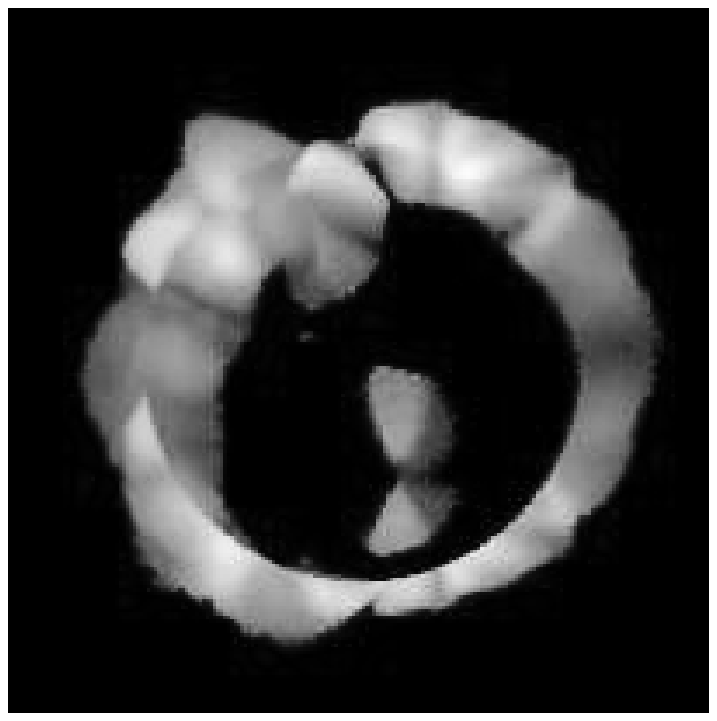,width=7cm}
\epsfig{file=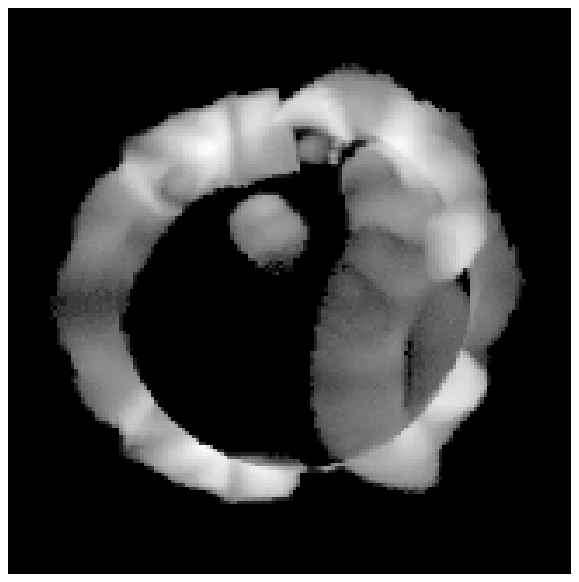, width=7cm}
\epsfig{file=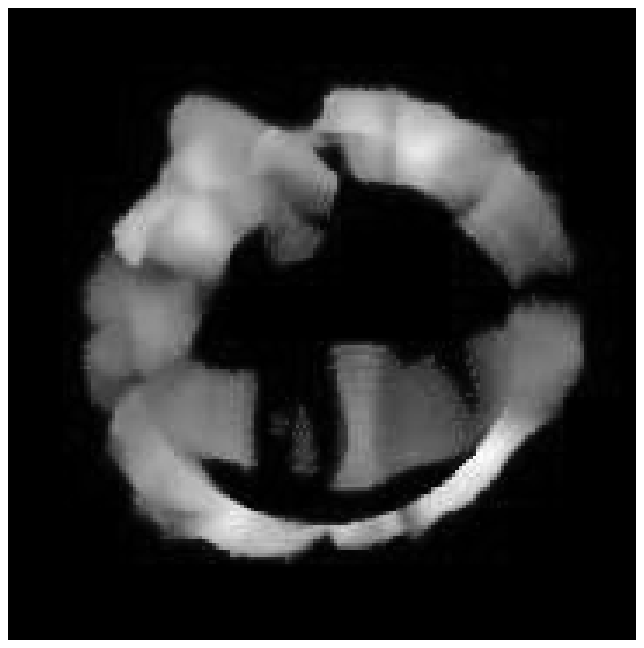, width=7cm}
\epsfig{file=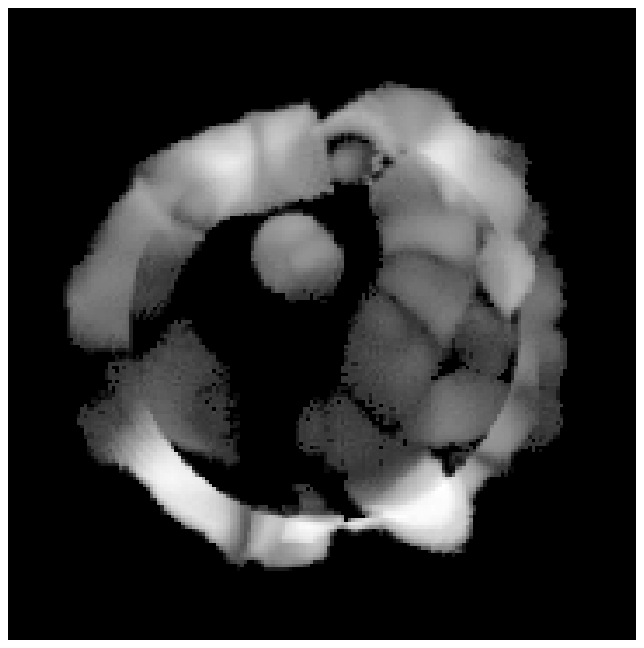,width=7cm}
\caption{Snapshots of the spot map and the X-ray models at rotation phases 0.3 (left column) and 0.8 (right column). 
{\em Top: } ``Snapshots'' of the surface spot distribution on AB Dor in 2002 December: grey and black represent unspotted and spotted regions of the photosphere respectively. The top left and top right images show the star at the maximum and minimum spottedness phases respectively.
{\em Middle row: } X-ray images generated from a magnetic field model in which the observed field is directly reflected into the obscured hemisphere (as the star has an inclination angle of 60$^{\circ}$). This is called model 'a'. {\em Bottom left and right: } In this X-ray model, the observed field is reflected into the obscured hemisphere {\em and its polarity is reversed}; this is called model 'b'. The X-ray emission pattern remains similar in the top hemisphere in both the 'a' and 'b' X-ray models, the main difference in model 'b' is that there are more field lines connecting the top and bottom hemispheres near the equator, thus increasing the X-ray emission contribution from this region. The X-ray lightcurves are computed by evaluating the total X-ray emission from X-ray images covering the full range of phases (see Figure\,7).
}
\label{fig:models}
\end{figure*}

One of the main objectives of this paper is to learn whether or not 
these magnetic field maps and X-ray observations can 
be used to build more realistic detailed X-ray models. 
As AB Dor is inclined at 60$^{\circ}$, we cannot observe the magnetic field in the obscured hemisphere. We model this ``missing'' field by reflecting the magnetic field in the observed hemisphere into the obscured hemisphere. Figure\,\ref{fig:models} shows the resulting X-ray models for two cases: when the observed field is simply reflected into the lower, obscured hemisphere; and when the observed field is reflected and its polarity is reversed in the lower hemisphere. As shown in Figure\,\ref{fig:models}, the different types of model do not affect the X-ray emission from the observed upper hemisphere; however, the reversed polarity model causes more X-ray emission to originate near the equator of the star as more field lines are introduced that connect the upper and lower hemispheres if the polarities are reversed filling in some areas. 

Two sets of models are generated following the above procedure. In group A: the field in the observed hemisphere is reflected into the lower hemisphere, and this lower hemisphere field is then offset relative to the field in the observed hemisphere by different amounts in longitude-space; X-ray lightcurves  are computed for each of these models and the rotational modulation evaluated.  In group B: the field is reflected, the polarity is reversed and this modified field is then displaced relative to the field in the upper hemisphere by different amounts in longitude; as with group A, X-ray lightcurves are computed for each model and rotational modulation measured. The observed level of rotational modulation in the lightcurve determined in Paper 1 is $\pm 12$\%. We verified that this level of rotational modulation is reliable by also constructing a X-ray lightcurve using just the flux in the strongest line profiles, hence reducing any contribution from  the continuum flux in the dataset. The resulting lightcurve also shows a consistent level of $\pm 12$\% rotational modulation superimposed with stochastic short-term variability. 

\subsection{X-ray lightcurves}

\begin{figure*}
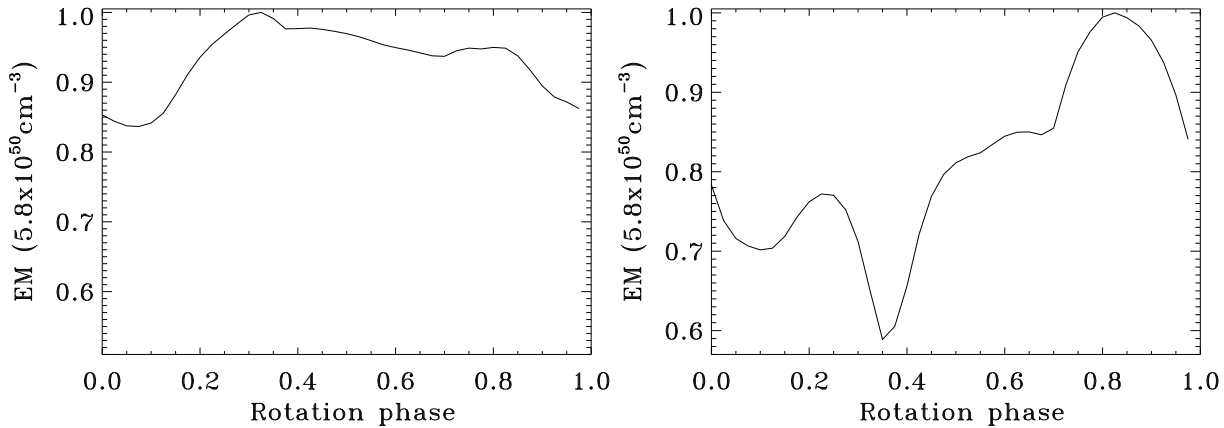

\epsfig{file=rotmod_2002.epsi,width=8cm}
\epsfig{file=rotmod_3215.epsi, width=8cm}
\caption{
{\em Left:} X-ray lightcurve for model 'a' in Figure\,\ref{fig:models}. 
{\em Right:} X-ray lightcurve for model 'b' in Figure\,\ref{fig:models}. 
Model 'b' has a much higher  level of X-ray rotational modulation than observed for AB Dor at this, or any other, epoch; therefore it is likely that the field in the obscured hemisphere is of the same polarity as that in the opposite hemisphere. }
\label{fig:model_lcurve}
\end{figure*}

Paper I uses a sine-curve fitting technique to quantify 
the rotational modulation in the {\em Chandra} X-ray lightcurve and measure $\pm$ 12\% rotational modulation from the mean level of the lightcurve. This is consistent with previous 
estimates of X-ray rotational modulation for AB Dor, with estimates varying between 
$\pm 5$--13\% (K\"urster et al. 1997; Brandt et al. 2001).

\begin{figure*}
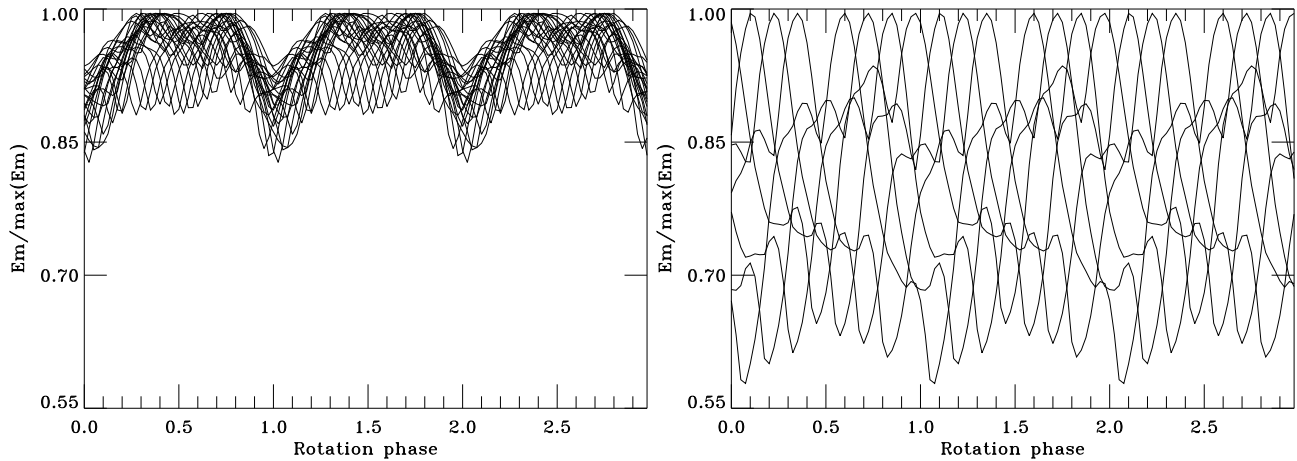

\epsfig{file=lightcurve_2002.epsi,width=8.5cm}
\epsfig{file=lightcurve_3215.epsi, width=8.5cm}
\caption{{\em Left: } X-ray lightcurves for the sets of models in which the  observed field is reflected into the obscured hemisphere. Introducing different offsets between the observed field and the reflected field changes the structure of the lightcurve but the level of rotational modulation remains the same ($\pm 8$\%).
{\em Right: }X-ray lightcurves for the sets  of models in which the observed field is reflected into the obscured hemisphere  and its polarity switched. As with the first set of models, different offsets between the two hemispheres affects the structure of the lightcurve but not the level of rotational modulation observed.}
\label{fig:model_lcurve_all}
\end{figure*}

Lightcurves generated from our two sets of models are shown in 
Figure\,\ref{fig:model_lcurve_all}. 
These plots confirm that the field in the obscured hemisphere would have a significant effect on the  observed  X-ray rotational modulation. 
The models in group A show rotational modulation of approximately $\pm 8$\%, while models in group B show much larger levels of X-ray rotational modulation ($\pm 21$\%).

We find that the detailed shape of the lightcurves can vary considerably in models with different phase offsets between the two hemispheres. However, the overall amount of rotational modulation in the models in each group does not vary significantly. These findings suggest that the detailed shape of an observed X-ray lightcurve would be hard to fit as it is strongly dependent on field that we cannot detect directly in ZDI surface magnetic field maps; we therefore concentrate  on understanding the family of models for which we can reconstruct the same level of X-ray rotational modulation as that observed in the {\em Chandra} dataset. 
The group A set of models provides the best fit to the level of rotational modulation observed in AB Dor, not only at this epoch but also at the previous epochs for which there have been long X-ray observations of the star (K\"urster et al. 1997). This suggests that 
the field in the obscured hemisphere is of the same polarity as that in the magnetic field map, unlike the solar case,
 where the leading polarities are reversed in opposite hemispheres.

\subsection{X-ray line profiles}

Using the coronal X-ray models shown in the previous section,
we produce simulated X-ray line profiles assuming the corona is optically thin and co-rotating with the star.  In  forming the X-ray images, we now take into account the Doppler shifts of  the photons emitted from different parts of the corona.  We assume the  X-ray line emission is at a single rest-frame wavelength (infinitely  sharp line emission) and by binning the emitted photons according to  their Doppler shifts we form X-ray channel maps.  Line profiles are then formed by integrating the total intensity in the individual channel  maps.

Forty line profiles are produced, evenly sampling a full rotation cycle. We phase-bin the line profiles to produce quarter-rotation-phase binned line profiles for  three different X-ray models: (a) with a source surface at 1.3\,R$_*$ and the magnetic field reflected into the obscured hemisphere, (b) with a source surface at 1.3\,R$_*$ with the reflected magnetic field in the obscured hemisphere possessing the reversed polarity and (c)  with the magnetic field reflected into the obscured hemisphere and a source surface at 3\,R$_*$. This last model has the same temperature and density as the other two models but with a higher EM ($\log {\rm EM[cm^{-3}]}>53$). 
The line profiles are then convolved with an instrumental profile corresponding to the 18.9\,\AA\ line in the LETG.
We measure the velocity shifts for these models using the same method as those found in Paper 1 (also shown in Figure 1), i.e. by fitting  Gaussian profiles in the data analysis package, {\em Sherpa}.  

\begin{figure}
\epsfig{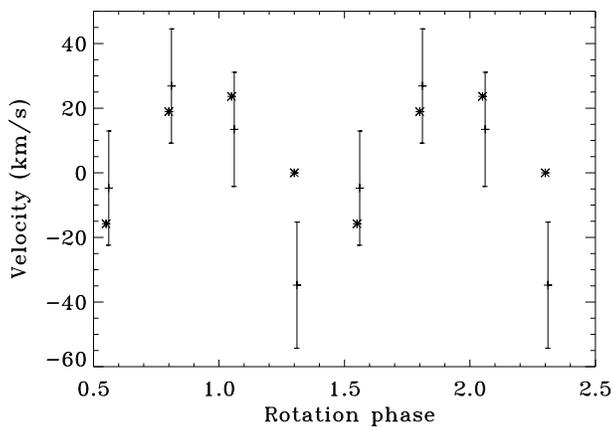}
\caption{The error bars and crosses show the velocity shifts measured in the line centroids of the observed LETG OVIII 18.9\AA\ profiles. 
The asterisks show the velocity shifts measured for our simulated profiles for the model that best reproduces the lightcurve modulation level observed. }
\label{fig:velshifts}
\end{figure}

The velocity shifts found in each model depend somewhat on the exact timing used, choosing a slightly different starting point can change the maximum size of the velocity shifts observed. However, the model (a), which was constructed by reflecting the magnetic field into the obscured hemisphere produces the largest velocity shifts (Figure~\ref{fig:velshifts}). Models (b) and (c), which possess reversed polarities and an extended source surface radius respectively, show significantly smaller velocity shifts than those measured in the {\em Chandra} spectra. This is probably because they posses more uniformly distributed X-ray emission that effectively fills in the centres of the line profiles and reduces the measured velocity shifts. 

As expected, increasing the spectral resolution does not affect the centroid measurements significantly  provided there is sufficient signal: increasing spectral resolution by a factor of four alters the positions of the line centroids by $\pm 1.5$\,km/s. 
Naturally, improving the time-resolution, and producing more phase bins over the rotation cycle would increase the sizes of the velocity shifts observed. If the line profiles were of higher S:N, then one would also observe larger velocity shifts in the {\em peaks} of the line profiles. 
With optical Doppler imaging the line profiles are inverted to obtain surface maps; 
these X-ray line profiles have significantly lower spectral and time resolution and so a similar inversion would  not be as effective here. These results suggest that  forward-modelling: creating X-ray models from surface field extrapolations and testing  these using the rotational modulation observed in both X-ray lightcurves and spectra is a more effective approach in the case of lower resolution spectra such as these.

\begin{figure}
\epsfig{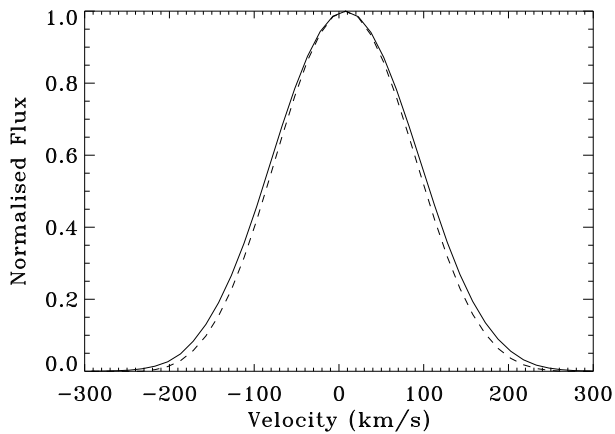}
\caption{Synthetic spectral line profiles produced by integrating the simulated X-ray line profiles from coronal models 'a' and 'c' over a full rotation cycle and convolving with the HEG instrumental profile corresponding to the HEG Fe XVII 15\AA\ line,  (FWHM$=0.009$\,\AA). The flux has been normalised by the maximum flux to enable easy comparison of both line profiles. The dashed line is for the compact X-ray model 'a' ($R_{\rm ss}=1.3$\,R$_*$) and the solid line is the extended model 'c' ($R_{\rm ss}=3$\,R$_*$). The instrumental resolution is insufficient to distinguish between X-ray coronae with different coronal extents reliably.}
\label{fig:spectotal}
\end{figure}

We use our models to test whether or not we can distinguish between X-ray coronae with source surfaces at distinctly different heights using X-ray spectral line profiles. In Paper 1, we use the  Fe\,XVII\,15.0\,\AA\ line profile from {\em Chandra} High Energy Grating (HEG) archive observations of AB Dor to measure the line profile width of the X-ray emission model and estimate that the bulk of the emitting corona does not extend beyond 1.3\,R$_*$.
We simulate this using our models by producing line profiles at fourty evenly spaced phases,  integrating these over the full rotation cycle, and convolving the integrated line profiles with the instrumental profile for the Fe\,XVII 15\,\AA\ line in the {\em Chandra} high energy grating (HEG) respectively.
However, as shown in Figure\,\ref{fig:spectotal}, by convolving the integrated spectral line with the instrumental profile of the HEG Fe\,XVII line, the current instrumental resolution is insufficient to distinguish between X-ray coronae that extend to 1.3\,R$_*$ and 3\,R$_*$. 
If the spectral resolution is improved by a factor of two, (i.e. if the instrumental profile has a FWHM$\sim 80$\,km/s) then the effect of extended X-ray emission would be visible as excess emission in the wings of the 3\,R$_*$ model, although very high S:N would be required to distinguish between different coronal extents  as the flux in the broadened wings is still only about 10\% of the peak flux in the line profile.  
The main evidence for AB Dor's compact  X-ray corona ($H\sim 1.4$\,R$_*$) therefore comes from the mixed polarities measured on its surface combined with the very high electron densities measured in the {\em Chandra} X-ray dataset.

\section{The extent and filling factor of the X-ray coronal model}

Measurements of pressure in AB Dor's corona are extremely high, over two orders
of magnitude larger than those found on the quiet Sun. Data from HST, EUVE, {\em Chandra},  and XMM-{\em Newton} indicate that  electron densities are similarly elevated in AB Dor, and other similarly magnetically active rapidly rotating cool stars.
$n_e \approx 2-3 \times 10^{12}$~cm$^{-3}$, at transition region
temperatures (about $3\times 10^4$~K)  (Brandt et al. 2001). 
At higher temperatures, around $3 \times 10^6$K, recent XMM-{\em Newton} results
indicate electron densities of $n_e \approx 3 \times 10^{10}$~cm$^{-3}$
({G\"udel} et al. 2001, Sanz Forcada et al. 2003, Testa et al. 2004). 
At diagnostics formed at even higher temperatures: between 6-10\,MK there is evidence for extremely high densities ($10^{12}<n_e<10^{13}$~cm$^{-3}$).
 Sanz-Forcada  et al. (2003) and Testa et al. (2004) argue that these very high densities indicate the presence of very compact structures, with heights of under 1$R_*$. 

\begin{figure}
\epsfig{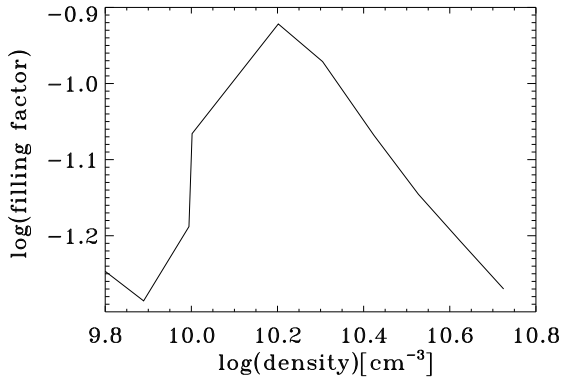}
\caption{The dependence of the X-ray corona filling factor on the coronal density for a coronal model with a $\log {\rm EM [cm^{-3}]}=52$ and a $T\sim10$\,MK. The filling factor is defined in the text. }
\label{fig:filling}
\end{figure}

The volume filling factor effectively measures the fraction of the coronal volume that is 
emitting in our X-ray model. 
The  volume filling factors
for our coronal models depend on the coronal density 
(Jardine et al. 2002), and are defined as follows:
\[f=\frac{\int n^2_e dV}{\frac{4}{3}\pi\left(R^3_s-R^3_*\right)n^2_e}.\]
As shown in Figure\,\ref{fig:filling}, $f\sim 0.11$ for our model that 
has a 10\,MK corona and a $\log n_e{\rm [cm^{-3}]}=10.3$. 
For higher densities, $\log n_e{\rm [cm^{-3}]} > 11$, 
the filling factor would decrease as the extent 
of the corona is smaller; for these models this would mean the X-ray lightcurves
would show more rotational modulation than we observed with the {\em Chandra}
dataset. Our models cannot reconstruct structures that are small enough to reproduce
these very high densities as they are extrapolated from the surface maps. 
The sizes of the smallest coronal structures that
can be modelled using this technique is constrained by the spatial resolution limit of the
surface maps ($\sim 3^{\circ}$).

\subsection{The effect of changing the coronal temperature on the sizes of X-ray emitting structures}
\begin{figure*}
\epsfig{file=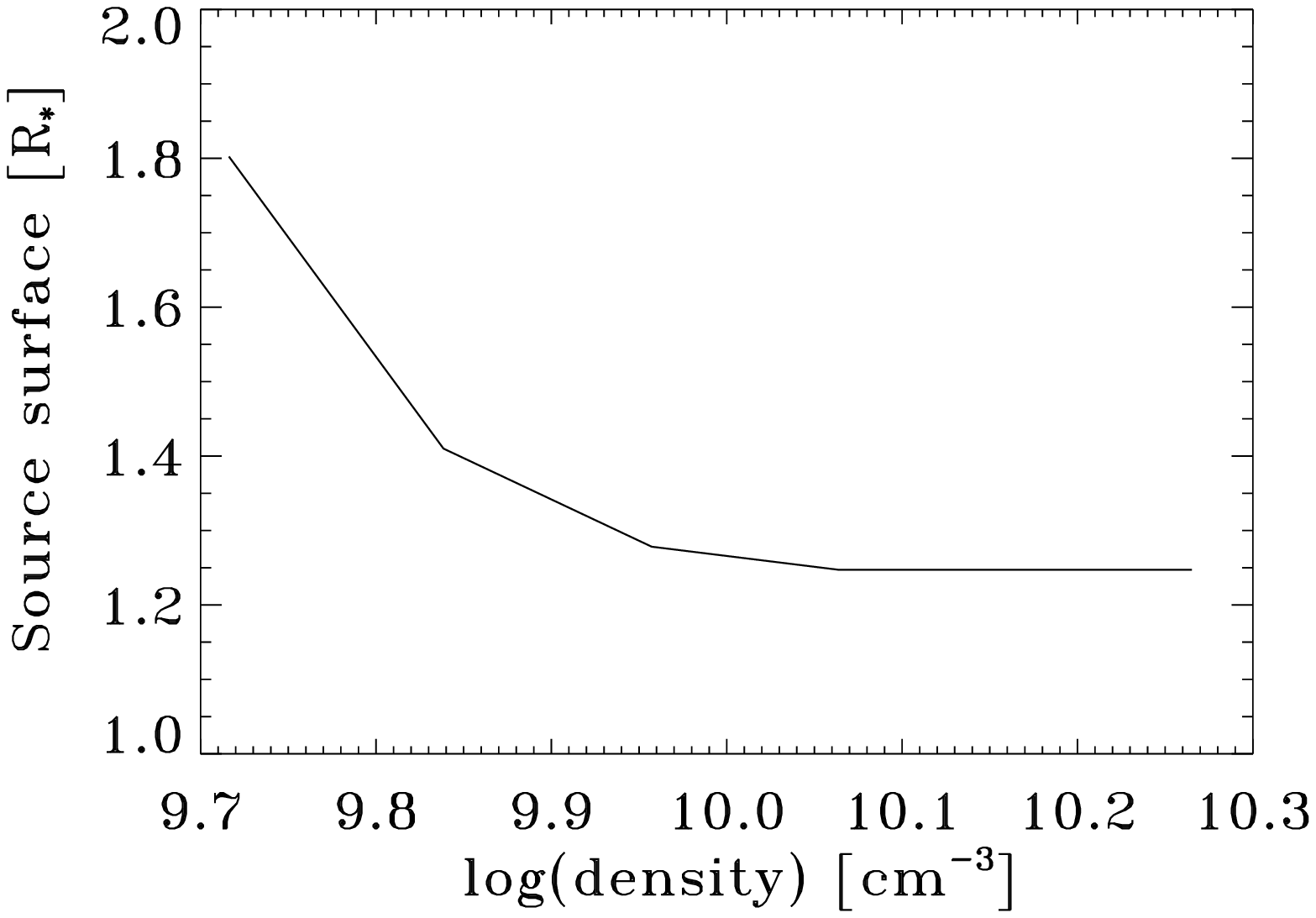, width=8.5cm}
\epsfig{file=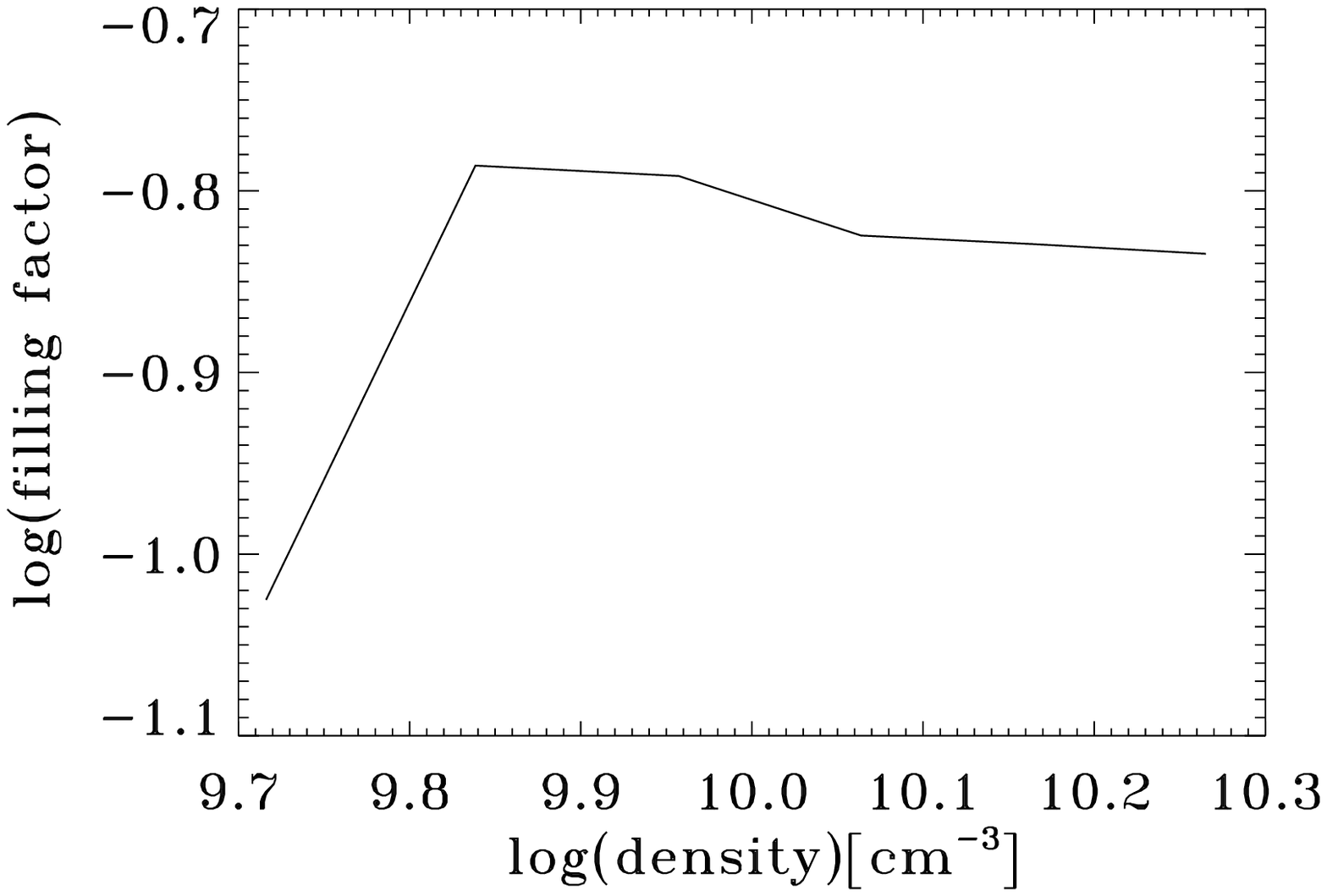, width=8.5cm}
\caption{A lower temperature model: assuming an isothermal corona with T=2.2\,MK and $\log {\rm (EM) [cm^{-3}]}=51.8$. {\em Left:} The relationship between the extent of the X-ray emitting corona and the coronal density, assuming the magnetic field map shown in Figure\,1 and the lower temperature coronal model properties. {\em Right: } The dependence of filling factor on $n_e$, also for the lower 2.2\,MK coronal model. The measured  density, $\log n_e{\rm [cm^{-3}]}=10.3$, suggests a very compact corona with a very similar source surface radius value to that estimated in the case of the 10\,MK corona,  $R_{\rm ss}\sim1.3$R$_*$ compared to 1.4$R_*$ for the hotter corona. The filling factor for this model would be slightly higher: $f=0.15$ compared to 0.11.}
\label{fig:model_rsslowt}
\end{figure*}

The stable peaks near $10$\,MK temperatures in EMDs
of cool stars may be caused by the presence of small, dense flaring structures that are always present on active cool stars (Cargill \& Klimchuk 2006). If this is the case, we may have overestimated the temperature of AB Dor's stable, quiescent, non-flaring corona. We investigate here the effect of lowering the coronal temperature on the coronal extent.  We have generated a series of models for a corona that is at 2.2\,MK, (i.e. the temperature determined from the density-sensitive line diagnostics at OVII 21\AA).  We use these to determine how changing the coronal temperature affects the extent of the X-ray emitting corona, $R_{\rm ss}$.
 The model that best fits this 2.2\,MK corona, and the measured density, 
$\log n_e {\rm [cm^{-3}]}\sim 10.3$,  will have a source surface radius of 1.3\,R$_*$.
It was not possible to find a solution for an EM of $\log {\rm EM [cm^{-3}]}=52$ and it had to be 
lowered to $\log {\rm EM [cm^{-3}]}=51.8$. This is consistent with what is observed,
as the EMDs of AB Dor obtained from EUV and X-ray spectra indicate that the EM of cooler 2.2\,MK
 plasma should indeed be lower than that of the 10\,MK plasma. 
The resulting X-ray model indicates that  changing the temperature by this amount does not significantly affect the extent of the X-ray corona as the high electron densities 
limit the size of the X-ray corona (Figure \ref{fig:model_rsslowt}).
The rotational modulation in the lightcurves for this more compact model increases 
to $\pm 15$\% in this lower temperature 2.2\,MK model (compared to $\pm 8$\% in the 
10\,MK model).  The filling factor increases slightly in this lower temperature model to 
$f\sim 0.15$.

\section{Prominences and plages in December 2002}
\protect\label{sect:proms}

The AB Dor \ha\ line is subject to variability caused primarily by bright surface plage regions and rapidly moving dark absorption features.  The latter are attributed to cool (8,000 to 10,000 K) clouds of neutral hydrogen in enforced co-rotation with the stellar surface (Collier Cameron and Robinson 1989a).  By measuring the rate at which the absorption features drift through the rotationally broadened \ha\ profile their distances from the stellar rotation axis can be measured. These are typically found to  be between 2 and 8 R$_*$ and therefore often above the Keplerian co-rotation radius of 2.7 R$_*$ (Cameron and Robinson 1989a, Cameron and Robinson 1989b, Donati  and Collier Cameron 1997, Donati et al. 1999).  These clouds must therefore be supported by magnetic loops extending high into the AB Dor corona, and are called `prominences' by analogy to those observed on the Sun.  Stellar prominences provide us with unique probes of the magnetic field topology at the highest levels in the stellar corona. 

\begin{figure*}
\epsfig{file=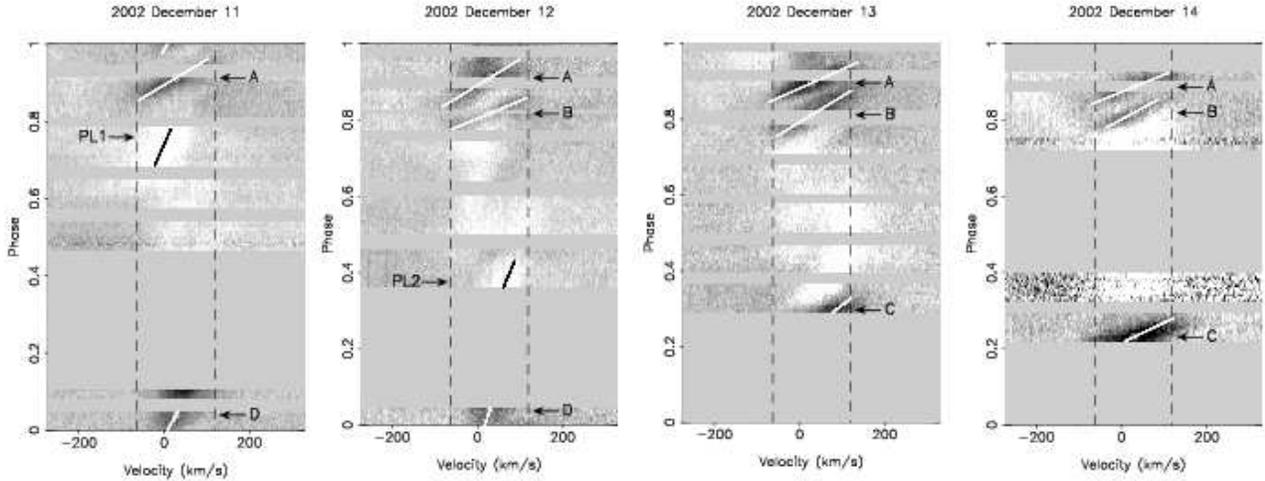,width=17cm}
\caption{
Raw \ha\ time series spectra for 2002 December 11, 12, 13 \& 14, with phase plotted against velocity.  Dashed black lines show the stellar \vsini\ limits.  The grey-scale runs from black at 0.80 times the local continuum level, to white at 1.05 times continuum.  The dark  absorption transients indicate the presence of cool prominences lying above the stellar surface. H$\alpha$ emission  traces the locations of bright plage regions as they transit the projected stellar disk. By measuring the rate of these transits the heights of the absorbing and emitting structures can be deduced. 
A mean spectrum has been subtracted from the entire time-series. For reference with Table 4 
 prominences are labelled A,B,C and D and two of the bright surface regions are labelled PL1 and PL2.}
\label{fig:promha}
\end{figure*}

In Figure\,\ref{fig:promha} we display the stacked \ha\ (time-series) spectra for each of the four nights of observations.  We approximate the background in the \ha\ line by creating a mean spectrum that is then subtracted from the entire time-series.  This leaves only time-varying features such as bright active regions (plage) and rapid absorption transients (prominences).  When examining Figure\,\ref{fig:promha} the most striking observation is the {\bf{lack}} of prominences seen at this epoch.  Despite the fact that almost 90\% of the rotation cycle is covered over four nights,  only four prominences are found.  This is substantially less than the average number of 6 to 8 prominences that are observed at most other epochs (Collier Cameron et al. 2003).  Furthermore, the distribution of prominences around the star also appears to be peculiar, as almost half of the star (between phases 0.3 to 0.8) does not appear to possess any prominences that transit the disk, and is covered in bright chromospheric emission from plage regions instead.  This is unusual for AB Dor, as prominences are typically distributed at a much wider range of longitudes. However, a similar asymmetric distribution has been reported on K3 dwarf Speedy Mic  (Dunstone et al. 2006a). Note that the longitudes that are effectively prominence-free (phases 0.3--0.8) correspond to the most magnetically-active longitude region in the surface spot and magnetic field maps (compare Figure\,\ref{fig:promha} with Figure\,\ref{fig:activelon}).

We use the simple technique of Gaussian-fitting (Collier Cameron 1989a) to track the \ha\ absorption signatures and measure the spatial and radial positions of the corresponding prominences.  Our best-fits are shown in Figure\,\ref{fig:promha} and our measurements of the phases and distances of these prominences are displayed in Table \ref{tab:results}.  Despite  prominences being detected at similar phases on successive nights their heights vary a lot   indicating that they may not be exactly the same structures from night to night. Instead they are stable locations in which prominences tend to reform quickly after being ejected. Further details and analysis of these \ha\ observations will be described by Dunstone et al. (in prep.). Prominences A, B and C are at distances of between 2--4\,R$_{*}$, and are typical of prominences found on AB Dor. Prominence D is quite different as it is at a high latitude and must be relatively close to the stellar surface to transit the disk: the distance from the rotation axis indicated by its drift rate is 0.91\,R$_{*}$ on December 11 and 0.52\,R$_{*}$ on December 12. This prominence first appears as an absorption feature in the \ha\ line at a velocity of -10\,\kms\ as opposed to the $\approx$-60 \kms\ of the other prominences.  
This confirms that the prominence starts to transit the  stellar disc at a high latitude ($\sim$\,60\degs).  Unfortunately as the  phase coverage is poor and the end of prominence D's transit is not observed our determination of its exact height is uncertain.  It is nevertheless unusual to find such a low lying, high latitude, prominence on AB Dor. 

In addition to the prominence absorption features we also track two bright plage regions,  labelled PL1 and PL2 in Figure \ref{fig:promha}.  Both of these, like prominence D, are found to have distances from the rotation axis less than 1 $R_{*}$.  If we assume that PL1 and PL2 are indeed active plage regions on the AB Dor surface then their latitudes are 44\degs\ and 49\degs\ respectively.  It is also interesting how near their phases of observation are to those of prominences B and C respectively.  This is similar to the observation by Jeffries (1993) that prominences on Speedy Mic were followed or preceded by strong emission transients.

Soft X-ray emission from the X-ray coronal active regions may be absorbed by these transiting stellar prominences, 
 (e.g., as seen in the X-ray {\em Einstein} lightcurve of the M5V, Proxima Centauri; Haisch et al. 1983). 
Now that the phases at which prominence material is transiting the stellar disc have been measured, we can compare these to the X-ray lightcurve. We only consider the parts of the rotation cycles over which there is an overlap between the X-ray and optical observations as prominences can form quickly, within 2$P_{\rm rot}$. Only prominences A and D are known to exist at the same time as the  {\em Chandra} observation. The possibility of soft X-ray emission being shadowed by cool prominence material, as observed on the Sun, was briefly discussed in Paper I.  Stellar prominences on rapidly rotating stars have higher hydrogen column densities than their solar namesakes. Collier Cameron et al. (1990) found a column density for AB Dor prominences of $N_{H}\approx10^{20}$ cm$^{-2}$, while more recently Dunstone et al. (2006b) found column densities in the range $N_{H}=$1  - 4$\mathrm{x} 10^{19}$ cm$^{-2}$ on \speedy.  This is an order of magnitude more than the $N_{H}\approx10^{18}$ cm$^{-2}$ found for solar prominences (e.g. Kucera 1998).

\begin{figure}
\begin{center}
\includegraphics[height=8cm,angle=270]{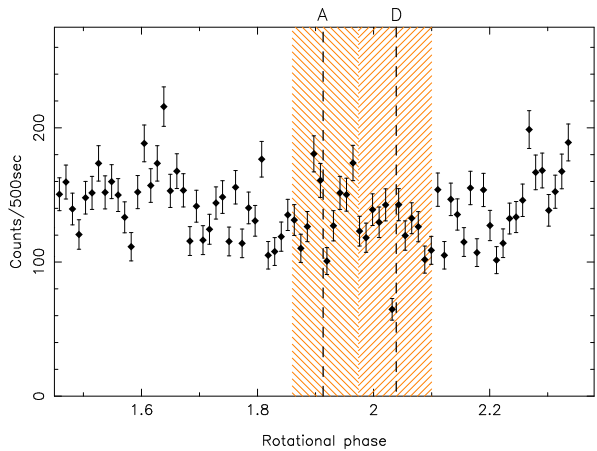}
\end{center}
\caption{EUV light curve obtained by summing up counts over the 50 - 130 \AA\ range, focussing on the phases over which there are simultaneous optical and X-ray observations.  The hashed regions show the phases at which different prominences are transitting the stellar disc, while the vertical black dashed lines show the phase of central meridian crossing.  Note the correlation between a rapid dip in the lightcurve at phase 2.03 and prominence D. }
\protect\label{fig:xraylcproms}
\end{figure}

The hydrogen continuum has an upper bound of 911\,\AA, below which the absorption cross-section decreases as $\lambda^{-3}$.  
The effect should therefore be strongest in the long wavelength region 
of our data (50--130\AA). This lightcurve is shown in Figure\,\ref{fig:xraylcproms} with the prominence positions overlayed. This plot focusses on the prominences that were observed
strictly simultaneously with the X-ray observations. 
The flux-weighted average wavelength of this lightcurve is 83.9\AA.  At this wavelength for hydrogen column densities in the range $10^{19}$ to $10^{20}$ cm$^{-2}$ we would expect a prominence to absorb between 5 and 45\% of EUV photons.  If the prominences also contain about 5\% He I, then the absorption increases to between 15--75\%. Furthermore if the prominences have a fragmented structure similar to solar prominences, then the local density and corresponding absorption could be higher. So there is reason to expect that a significant reduction in EUV flux would be observed if a prominence eclipsed an EUV emitting region of the corona.
 In contrast, approximately 1--2\% of the hard X-ray photons from the 1-50\AA\ lightcurve are expected to be absorbed; this would not be detectable due to the intrinsic variability.

It should be noted that the only data that are observed simultaneously at both X-ray and optical wavelengths lie between phases 1.5--2.1, i.e. during the December 11 spectral time-series. We include the X-ray observations of the previous AB Dor rotation. Initial investigation using 1\,ksec bins (Figure \,1a) revealed a possible weak correlation between the location of prominence D and a sharp dip in the X-ray light curve.  To establish the duration of this event more accurately we re-binned the light curve into 500\,s bins to produce Figure\,\ref{fig:xraylcproms}.  The dip is stronger but only apparent in one data point; we therefore increased the sampling further to produce a lightcurve with 250\,s bins and found that this point split into two equal fluxes, which lay below the other datapoints. So it appears that for 500\,s the EUV flux decreases considerably and it does so at a phase that corresponds closely to that of the central meridian crossing of prominence D.  The phase difference is $\Delta\phi=0.007$ (or $\approx300$\,s), which is the same as the exposure time of each \ha\ spectrum.

In order to establish the strength of this 500\,s dip and its significance we need to compare it to a reference flux level to mark the unmodulated emission level of the lightcurve. We measure this three ways: (a) the local average (average of the counts of the four data points immediately to either side of the dip), (b) the optimal average flux of the whole light curve,  and  (c) the minimum background level (found by taking an iterative 10$\sigma$ clip of all points above the optimal average level). The local and optimal average flux level are similar: 134 and 139 counts respectively; the magnitude of the dip is therefore between 52--54\%  corresponding to a significance of 8--9$\sigma$. The minimum background level is 114 counts, indicating a dip of 43\% and a significance of 6$\sigma$. Note that the significance of the dip discussed here only corresponds to the error associated with the flux measurement; there is also added uncertainty introduced by the intrinsic short-term variability observed in the {\em Chandra} lightcurves. 
 The short wavelength (1--50\AA) lightcurve shows no significant dip at the corresponding phase (see Figure\,1a). This lightcurve shows variations consistent with the intrinsic variability of the star; with variability around this phase being between 1--2$\sigma$ 
  (depending on exactly how the mean level of emission is  defined).
  This is consistent with the scenario of a cool prominence-type structure causing 
 the significant dip observed in the EUV lightcurve  as the effect of this type of cool structure 
 on the 1--50\AA\ lightcurve is expected to be much smaller (between 1--2\% compared to 
 15--75\% in the EUV lightcurve).

\begin{table*}
\caption{The results of the prominence tracking analysis displayed with phases of meridian crossing and calculated distances from the rotation axis ($\frac{{\varpi}}{R_*}$). }
\protect\label{tab:results}
\begin{center}
\begin{tabular}{cccccccccccc}
\hline
\multicolumn{1}{c}{Feature}     & \multicolumn{2}{c}{\it{December 11}} & \multicolumn{2}{c}{\it{December 12}}   & \multicolumn{2}{c}{\it{December 13}}  & \multicolumn{2}{c}{\it{December 14}}\\
& Phase &  $\frac{{\varpi}}{R_*}$ &     Phase & $\frac{{\varpi}}{R_*}$ & Phase & $\frac{{\varpi}}{R_*}$&  Phase & $\frac{{\varpi}}{R_*}$ \\

\hline
A               & 0.913 & $2.64\pm0.05$ & 0.910 & $2.55\pm0.03$ & 0.896 & $3.48\pm0.05$ & 0.887 & $3.84\pm0.12$\\
B               &       &               & 0.817 & $3.74\pm0.04$ & 0.811 & $2.55\pm0.03$ & 0.819 & $2.88\pm0.03$\\
C               &       &               &       &               & 0.247 & $2.00\pm0.04$ & 0.231 & $3.31\pm0.05$\\
D               & 0.039 & $0.91\pm0.03$ & 0.035 & $0.52\pm0.19$ &       &               &       &               \\
Plage 1         & 0.809 & $0.72\pm0.02$ &       &               &       &               &       & \\
Plage 2         &       &               & 0.274 & $0.66\pm0.02$ &       &               &       & \\
\hline
\end{tabular}
\end{center}
\end{table*}

\section{Discussion}
\subsection{Correlation between active longitudes at different types of magnetic activity}
There is a significant level of correlation in the positions of active longitudes when comparing the surface magnetic field maps with spot maps as well as when comparing magnetic field maps or spot filling factors with the X-ray lightcurve. Some agreement has been observed between the positions of active regions as traced in surface spot and magnetic field maps of AB Dor at previous epochs, although  this is not found in similar studies of other systems (e.g., K0V, LQ Hya, or the primary component of the RSCVn binary, HR1099) (see Donati et al. 2003). 
The correlation may not be so apparent in studies of these other systems as they are more slowly rotating and it is therefore more difficult to obtain the same degree of phase coverage and hence detail in their surface maps. Further analysis of detailed surface magnetic field maps of stars at different epochs would be necessary to establish how common this degree of correlation really is. We have carried out a similar analysis of  spot and magnetic field maps of AB Dor obtained at previous epochs (1998 January, 1996 December and 1995 December) and do find some correlation between  spot and magnetic field active longitudes although at a much weaker level than at 2002 December (the epoch analysed in this paper). Hence, it appears that  2002 December is an unusual epoch for AB Dor in two respects: (a) half of the star, between phases $\sim 0.1$--$0.5$, appears to be covered in stronger magnetic fields, increased spot coverage, and stronger chromospheric emission compared to the other half; and (b) cool H$\alpha$ absorbing prominences are not distributed across all longitudes as in previous epochs, rather they appear to be restricted to the less active phases (see Figure\,\ref{fig:promha}). At this epoch there appears to be a clear link between activity at all observed atmospheric levels. 

This is the first dataset for which surface spot, magnetic field maps have been acquired contemporaneously with X-ray observations and we clearly find significant correlation between both surface active longitudes (as traced by the spot and magnetic field maps) and the coronal active longitudes (as traced by the X-ray lightcurve). This correlation would, of course, have been obscured if there had been any large flares on AB Dor over the course of the X-ray observation. We investigate what level of correspondence we can realistically expect between surface and X-ray active longitudes using coronal models based on extrapolations of the surface magnetic field map of AB Dor in Section\,4.1 
and find that the unobserved field in the obscured hemisphere can have a significant impact on the shape of the X-ray lightcurve (see Figures 7 and 8). Hence, we cannot accurately reproduce the exact shape of the observed X-ray lightcurve without knowing the type of field in the obscured hemisphere.  Our results suggest that one would expect significant correlation between X-ray emission and optical variability caused by dark starspots in active main sequence stars.

\subsection{Sizes of X-ray emitting structures compared to other coronal features}

We have found that the size of the X-ray corona must be very small, with a height that is not greater than 0.4\,R$_*$. 
The main factors that limit the size of the X-ray loops are (a) high electron density, $n_e \sim 2\times 10^{10}$\,cm$^{-3}$, and (b) the complex multi-polar surface magnetic field distribution that is recovered using ZDI.
Sanz-Forcada et al. (2003) estimate that the 2\,MK corona can extend to 1.3\,R$_*$, 
while the 10\,MK corona would extend out to 0.24\,R$_*$. 
However, their models assume a global dipole, for which the magnetic field strength decays more slowly with height, $r$, 
$B \sim r^{-3}$; our ZDI maps of AB Dor clearly show a complex multi-polar field distribution; which decays much more quickly with height. As X-ray loops can only be supported when the magnetic pressure exceeds the electron pressure, we 
find that lowering the temperature to 2\,MK does not change the height of the source surface significantly as a compact corona is necessitated by the high $n_e$ value measured using density-sensitive diagnostics in  the {\em Chandra} X-ray data. 
Our estimate of the coronal  extent, between 0.3--0.4\,R$_*$, is consistent with the 0.3\,R$_*$ flaring loop height determined from a large  X-ray flare observed on AB Dor by Maggio et al. (2000). 

Some $n_e$ measurements of the hotter, 10\,MK plasma in AB Dor and other active cool stars indicate densities that are even higher than that measured at 2\,MK  with $12<\log n_e {\rm [cm^{-3}]}<13$ values (e.g., Sanz-Forcada et al. 2002, Testa et al. 2004). This would imply that a component of the corona at 10\,MK is extremely compact, with a height, $H \ll 0.1$\,R$_*$. 

\subsection{Complex multi-thermal coronae}

We present tentative evidence for EUV absorption due to the presence of a cool prominence transiting near the stellar pole. This is detected as a strong dip in the EUV lightcurve near the meridian crossing of prominence; no corresponding dip is observed in the short wavelength (1--50\AA) lightcurve. The short duration of the eclipse would suggest that both the prominence and the EUV emitting region are spatially narrow in longitude.  Given that the observed dip is approximately 50 \% of the entire EUV flux coming from AB Dor this would suggest that the prominence core is very dense and narrow and at least one dominant EUV emitting region is highly localised and near the pole.  This would then indicate that at least half of the EUV flux from AB Dor's corona originates near the stellar surface and at high latitudes (near the stellar pole) in fewer and more concentrated emitting regions than in our X-ray models. This region would contribute  to the unmodulated component of emission in the X-ray and EUV lightcurves. However,  as there is considerable intrinsic variability due to flaring, and the distance and position of the prominence could not be tracked fully, we cannot definitively measure the sizes of the obscuring and EUV emitting structures involved. In the future more simultaneous  X-ray and optical observations would allow us to better understand the physical properties of the prominences and the sizes and distribution of EUV emitting regions.

We conclude that the corona is likely to be made up of different types of structures, ranging from extremely hot, dense (presumably flaring) plasma very near the surface, to a less dense corona at a temperature of a few MK extending out to 0.3--0.4\,R$_*$, leading finally to an extended corona with a height that may reach the Keplerian co-rotation radius ($\sim$2.7\,R$_*$). This extended corona could support much cooler, low density material that would not be detectable in X-rays (e.g., cool prominences with temperatures of between 8000-9000\,K). While the models presented here are simple X-ray models, allowing only one coronal temperature and density; we have demonstrated that these models are useful in estimating the global properties of stellar coronae and can reproduce key characteristics of the X-ray observations: the densities, dominant temperatures and rotational modulation in the X-ray lightcurves and spectra. Our models show that the bulk of X-ray emission observed from the star is unmodulated due to a quite large component that is located in the high latitude regions and at the stellar pole (see Figure\,\ref{fig:models} for X-ray images). In the future, we will build more complicated coronal models, allowing a range of temperatures and densities based on extrapolations of surface magnetic field maps. Future multi-wavelength campaigns, co-ordinating X-ray, UV observations with Doppler imaging and Zeeman Doppler imaging studies of other active cool star systems will enable us to understand how representative these models are of cool stars in general.

\subsection{Summary}

\begin{itemize}
\item  We have obtained contemporaneous X-ray and optical data probing magnetic activity at the surface and corona of AB Dor at the same time. 
\item We find significant correlation between the positions of active longitudes in all our activity diagnostics as determined from the X-ray lightcurve and the surface spot and magnetic field maps. At this epoch, AB Dor appears to possess one very large active longitude region that displayed enhanced spot coverage, magnetic fields and chromospheric emission. 
\item We produced detailed three-dimensional models of the quiescent X-ray corona of AB Dor by extrapolating surface magnetic field maps of the system; using coronal densities and temperatures determined from X-ray data of the system. Our models indicate that the X-ray emitting corona must be concentrated very close to the surface of AB Dor, with a height, $H\sim0.3$--0.4R$_*$.
\item The X-ray models suggest that the ``missing'' flux in the obscured hemisphere of the star needs to be of the same polarity as that in the observed hemisphere in order to reproduce the same level of rotational modulation observed in the X-ray lightcurves and spectra. 
\end{itemize}

 \section*{Acknowledgments}
 GAJH gratefully acknowledges the support of an ESA Fellowship and PPARC funds towards
 this work. NSB was supported by the Chandra X-ray Center through NASA contract 
 NAS8-03060 to the Smithsonian Astrophysical Observatory.
 Support for this work was also provided by NASA through Chandra Award Number GO3-4022X 
 issued by the Chandra X-ray Observatory Center, which is operated by the Smithsonian 
 Astrophysical Observatory for and on behalf of NASA.  
 We are also grateful to the referee, Manuel G\"udel, for his careful reading of the manuscript and
 helpful suggestions that have improved the quality of this paper.

\label{lastpage}

\end{document}